\documentclass [11pt,letterpaper]{JHEP3}
\usepackage{graphicx}
\usepackage{epsfig}
\usepackage{epic}       
\usepackage{cite}	
\usepackage{psfrag}     
\usepackage{bm}		
\usepackage{verbatim}   
\advance\parskip 1.0pt plus 1.0pt minus 2.0pt	
\def\coeff#1#2{{\textstyle {\frac {#1}{#2}}}}
\def\half{\coeff 12}
\def\l{\ell}
\def\j{{\bf j}}
\def\r{{\bf r}}
\def\d{{\rm d}}
\def\tr{{\rm tr}}
\def\Re{{\rm Re}\,}
\def\Nc{N_{\rm c}}
\def\Ns{N_{\rm s}}
\def\Nf{N_{\rm f}}
\def\Z{{\mathbb Z}}
\def\dlangle{\langle\!\langle}
\def\drangle{\rangle\!\rangle}

\def\betatilde{\widetilde\beta}

\newcommand{\drawsquare}[2]{\hbox{%
\rule{#2pt}{#1pt}\hskip-#2pt
\rule{#1pt}{#2pt}\hskip-#1pt
\rule[#1pt]{#1pt}{#2pt}}\rule[#1pt]{#2pt}{#2pt}\hskip-#2pt
\rule{#2pt}{#1pt}}
\newcommand{\Yfund}{\raisebox{-.5pt}{\drawsquare{6.5}{0.4}}}

\preprint {UW/PT 03--25\\INT-PUB 03--18}
\title
    {%
    Non-perturbative equivalences among large $\bm \Nc$ gauge theories
    with adjoint and bifundamental matter fields
    }%
\author
    {%
    Pavel Kovtun, Mithat \"Unsal, and Laurence G.~Yaffe
    \\Department of Physics
    \\University of Washington
    \\Seattle, Washington 98195--1560
    \\Emails: 
    \parbox[t]{2in}{\email {pkovtun@phys.washington.edu},\\
		    \email {mithat@phys.washington.edu},\\
		    \email {yaffe@phys.washington.edu}}
    }%

\abstract
    {%
    We prove an equivalence, in the large $\Nc$ limit,
    between certain $U(\Nc)$ gauge theories containing adjoint representation
    matter fields and their orbifold projections.
    Lattice regularization is used to provide
    a non-perturbative definition of these theories;
    our proof applies in the strong coupling,
    large mass phase of the theories.
    Equivalence is demonstrated by constructing and comparing
    the loop equations for a parent theory and its orbifold projections.
    Loop equations for both expectation values of single-trace observables,
    and for connected correlators of such observables, are considered;
    hence the demonstrated non-perturbative equivalence applies
    to the large $\Nc$ limits of both string tensions and particle spectra.
    }%


\begin {document}

\section {Introduction}

Various examples are known in which two gauge theories with $\Nc$ colors,
which differ for all finite $\Nc$,
become indistinguishable in the $\Nc\to\infty$ limit.
Such equivalences
include lattice Yang-Mills theories with mixed fundamental-adjoint
representation actions,
whose leading large $\Nc$ limits coincide with those of
pure fundamental representation actions (provided
one suitably modifies the value of the lattice gauge coupling)
\cite{Makeenko-Polikarpov,Samuel}.
Another example is the
volume independence of large $\Nc$ gauge theories.
This is often referred to as Eguchi-Kawai reduction \cite{Eguchi-Kawai};
see Refs.~\cite{Das,Neuberger1,Neuberger2} for more recent discussions.
Such large $\Nc$ equivalences can also involve theories with
differing gauge groups.
As a trivial example,
the existence of the $\Nc\to\infty$ limit in
$U(\Nc)$ pure gauge theories implies that
Yang-Mills theories with gauge groups $U(\Nc)$ and $U(k\Nc)$
(for any positive integer $k$)
and coinciding 't Hooft couplings
are indistinguishable as $\Nc\to\infty$.

Recently, possible large $\Nc$ equivalences
between pairs of theories related by so-called
``orbifold'' projections have received attention
\cite{Douglas-Moore,Kachru-Silverstein, Lawrence-Nekrasov-Vafa,
Bershadsky-Kakushadze-Vafa, Bershadsky-Johansen, Schmaltz, Erlich-Naqvi, 
Strassler, Gorsky-Shifman, Dijkgraaf-Neitzke-Vafa, Tong}.
In this context, orbifold projection is a technique
for constructing a ``daughter'' theory, starting from some
``parent'' theory, by retaining only those fields which
are invariant under a discrete symmetry group of the parent theory.%
\footnote
    {
    The name ``orbifold'' comes from string theory,
    where daughter theories of this type originate
    as low-energy world-volume descriptions of
    $D$-branes on space-time orbifolds
    \cite{Douglas-Moore,Kachru-Silverstein,
    Lawrence-Nekrasov-Vafa,Bershadsky-Kakushadze-Vafa}.
    }
The basis for a possible large $\Nc$
equivalence between parent and daughter orbifold theories comes from 
the fact that in the large-$\Nc$ limit of ordinary perturbation theory,
planar graphs of the orbifold theory 
exactly coincide with planar graphs
of the original theory, up to a simple
rescaling of the gauge coupling constant
\cite{Bershadsky-Johansen}.
Because perturbation theory is only an asymptotic expansion,
coinciding perturbative expansions do not imply that two theories
must be equivalent.
In particular,
in asymptotically free theories
the mass spectrum is purely non-perturbative,
so coinciding perturbative expansions do not, by themselves,
imply that parent and daughter orbifold theories have identical
particle spectra.
However, the existence of a perturbative equivalence between parent and
daughter theories does make it natural to ask whether the large $\Nc$
equivalence is valid non-perturbatively.
If true, there are a variety of interesting consequences
\cite{Schmaltz,Strassler}.
For example, the fact that supersymmetric
theories may have non-supersymmetric 
orbifolds would imply that certain non-supersymmetric theories
must develop an accidental boson-fermion degeneracy
in part of their mass spectrum as $\Nc \to \infty$.

To date, no non-perturbative proof of large $\Nc$
equivalence between parent and daughter orbifold theories has been given.%
\footnote
    {
    Excluding the case of pure Yang-Mills theories,
    where large $\Nc$ equivalence under orbifold projection
    is nothing more than a repackaged form of the above-mentioned
    equivalence between $U(\Nc)$ and $U(k \Nc)$ Yang-Mills theories 
    in the $\Nc\to\infty$ limit.
    }
Several tests have been proposed to check whether
the equivalence might hold non-perturbatively,
both for supersymmetric \cite{Erlich-Naqvi},
and non-supersymmetric \cite{Gorsky-Shifman,Dijkgraaf-Neitzke-Vafa,Tong}
daughter theories.
Evidence consistent with a possible non-perturbative
large $\Nc$ equivalence has come from
comparison of the holomorphic couplings in
parent and daughter theories \cite{Erlich-Naqvi},
as well as from a matrix model analysis
of low-energy effective actions \cite{Dijkgraaf-Neitzke-Vafa}.
Various results \cite{Kachru-Silverstein,Lawrence-Nekrasov-Vafa}
on conformal field theories 
obtained from AdS/CFT duality
(which is widely believed, but unproven)
are also consistent with a non-perturbative large $\Nc$ equivalence.
 
However, evidence of large $\Nc$ \emph {inequivalence}
between certain parent and daughter orbifold theories
has also been found.
In particular, 
a mismatch between the number
of instanton zero modes in small volume was found
in Ref.~\cite{Gorsky-Shifman}, and
in Ref.~\cite{Tong} it was argued that a compactified
orbifold theory, unlike its parent, undergoes a phase 
transition at a non-zero value of the compactification radius.
These previously considered examples all involve cases where
the parent theory is supersymmetric, and no issues involving
non-perturbative regularization have been addressed.
Beyond these specific examples, it can only be said that
it is not yet clear under what circumstances
a non-perturbative equivalence does, or does not, hold.

In this paper, we will present a proof of the
non-perturbative equivalence between the large $\Nc$ limits
of a simple class of parent gauge theories and their orbifold projections.
Specifically, we will consider $U(\Nc)$ gauge theories containing
either scalar or fermion matter fields (or both) transforming in
the adjoint representation of the gauge group.
In order to have a rigorous basis for making non-perturbative arguments,
we will use a lattice formulation of our theories.
Physical observables of interest, including the mass spectrum
of the theory, can be extracted from correlation functions of
Wilson loops, possibly decorated with insertions of adjoint
representation matter fields.
These correlation functions obey a closed set of loop equations
in the large-$\Nc$ limit.%
\footnote
    {
    In the case of pure gauge theories, these loop equations
    are sometimes called the Migdal-Makeenko equations \cite {MM}.
    }
Our strategy will entail:
({\em i}) 
    showing that
    the large $\Nc$ loop equations of parent and daughter orbifold theories,
    for the relevant class of observables,
    coincide after trivial rescaling of coupling constants,
and ({\em ii}) 
    arguing that this system of loop equations can,
    at least in the phase of the theory continuously connected
    to strong coupling and large mass,
    uniquely determine the resulting correlation functions.

In other words, we will argue that comparison of large $\Nc$ loop equations
can, under appropriate conditions, provide a sufficient means for determining
when two theories have coinciding large $\Nc$ limits.
This idea is not new; essentially the same strategy has previously been
used in discussions of fundamental-adjoint universality
\cite{Makeenko-Polikarpov},
and the Eguchi-Kawai reduction \cite{Eguchi-Kawai}.
Our argument for the unique reconstruction of correlation functions
based on their loop equations will be completely rigorous in the
phase of the theory which is continuously connected to strong coupling
and large mass (for the matter fields).
The extent to which one can uniquely reconstruct correlation functions
from their loop equations more generally is discussed further in
Section~\ref{sec:Pure gauge}.

The paper is organized as follows.
In Section~\ref{sec:Pure gauge} 
we establish our notation and give
a self-contained review of loop equations in pure $U(\Nc)$ gauge theory.
This includes both loop equations for expectation values of
single Wilson loops, as well as the extension to multi-loop
connected correlators.
In this section we also discuss the reconstruction
of correlation functions from their loop equations.
The extension to gauge theories with matter fields in the adjoint 
representation is discussed in Section~\ref{sec:Adjoint matter}.
A key ingredient is the introduction of an ``extended'' higher-dimensional
lattice (with one new dimension for each matter field flavor)
in such a way that Wilson loops decorated with arbitrary insertions
of adjoint matter fields become isomorphic with simple loops on the
extended lattice.
This will allow us to formulate loop equations for theories containing
adjoint matter fields in a compact and elegant form which closely
mimics the loop equations of pure gauge theories.
In Section~\ref{sec:Orbifolds} we discuss orbifold projections
of $U(\Nc)$ gauge theories with adjoint matter fields.
In order to be self-contained, and establish notation,
we first review what is meant by an orbifold projection,
and then derive the loop equations in orbifold projected theories.
We observe that the loop equations are exactly the same in the
original theory and its orbifold projections provided that
(\emph{i}) observables in the two theories are appropriately identified,
(\emph{ii}) coupling constants of the two theories are suitably scaled,
(\emph{iii}) global symmetries used to define the orbifold projection
are not spontaneously broken in the parent theory, and
(\emph{iv}) global symmetries of the daughter theory which cyclically permute
equivalent gauge group factors are also not spontaneously broken.
Section~\ref{sec:Discussion} discusses implications of this
correspondence of loop equations between parent and daughter theories.
At least in the strong-coupling/large-mass phase of both theories,
we argue that this correspondence constitutes a non-perturbative proof of
the large $\Nc$ equivalence of parent and daughter theories.
Possible generalizations and extensions are also briefly discussed.
For simplicity of presentation, only correlation functions of bosonic
observables are considered in Sections \ref{sec:Adjoint matter}
and \ref {sec:Orbifolds}, but the extension to fermionic observables
is sketched in Appendix \ref{sec:Fermionic observables}.
Appendix \ref {sec:Iteration} describes
(in a more pedagogical manner than in the main text)
how one may see the presence of confinement and a mass gap
in the iterative strong-coupling solution of the loop equations.

\section{Pure gauge theory}
\label{sec:Pure gauge}

\subsection {Definitions}

Let $\Lambda$ denote a $d$-dimensional (Euclidean spacetime) lattice,
which may be either infinite or finite in extent.
To define a $U(\Nc)$ lattice gauge theory,
one associates a unitary matrix $u[\l] \in U(\Nc)$
with every directed link $\l \in \Lambda$ of the lattice.
Links (and plaquettes, {\em etc.}) are oriented;
we will use $\bar \l$ to denote the opposite orientation
of link $\l$, and $u[\bar\l] \equiv u[\l]^\dagger$.
Lattice Yang-Mills theory may be defined by the probability measure
\begin {equation}
    d\mu \equiv \frac {e^S}Z \; d\mu_0 \,,
\label {eq:dmu}
\end {equation}
where
\begin {equation}
    d\mu_0 \equiv {\prod_{\l\in \Lambda}}' \> du[\l] 
\label {eq:dmu0-YM}
\end {equation}
denotes the product of Haar measure for every (positively oriented)
link of the lattice,
\begin {equation}
    Z \equiv \int d\mu_0 \> e^S
\label {eq:Z-YM}
\end {equation}
is the usual the partition function, and
\begin {equation}
    S \equiv \beta \> {\sum_{p \in \Lambda}}' \> \Re \, \tr \, u[\partial p]
\label {eq:S0-YM}
\end {equation}
is the standard Wilson action involving a sum over all
(positively oriented) plaquettes in the lattice.
Here, $\partial p$ denotes the boundary of plaquette $p$ and
$u[\partial p]$ is the ordered product
of link variables around this plaquette boundary.
The prime on the product over links in the measure (\ref {eq:dmu0-YM}),
and on the sum over plaquettes in the action (\ref {eq:S0-YM}),
are indicators that only positively oriented links or plaquettes,
respectively, are to be included.
Subsequent unprimed sums over (various sets of) links or plaquettes
should be understood as not having this restriction.
The form (\ref {eq:S0-YM}), in which every plaquette contributes equally,
is appropriate for regular, isotropic lattices.
More generally, the contribution of a given plaquette may depend
on its orientation (or location), in which case
\begin {equation}
    S \equiv {\sum_{p \in \Lambda}}' \>
    \beta_p \> \Re\, \tr \, u[\partial p] \,,
\label {eq:S-YM}
\end {equation}
with $\beta_p$ some specified weight associated with every plaquette.
We will never introduce a lattice spacing explicitly;
all dimensionful couplings should be understood as measured in lattice units.

Wilson loops are the basic observables of the theory.
For any (directed) closed loop $C$ contained in the lattice,
let $u[C]$ denote the ordered product of link variables around
the loop $C$ (starting from an arbitrarily chosen site on the loop),
and define
\begin {equation}
    W[C] \equiv \frac 1\Nc \, \tr \, u[C] \,.
\label {eq:W[C]}
\end {equation}
(The factor of $1/\Nc$ is included so that expectation values
of Wilson loops have finite, non-trivial large-$\Nc$ limits.)


\subsection {Loop equations}

Loop equations are, in effect, Schwinger-Dyson equations for
the expectation values of Wilson loops (or their products)
\cite{MM,Forster,Eguchi,Weingarten,Wadia}.
To generate these loop equations, it is convenient to define
operators $\delta^A_\l$ which vary individual link variables.
Specifically,
\begin {equation}
    \delta^A_\l \> u[\l'] \equiv \delta_{\l\l'} \, t^A \, u[\l] \,,
\label {eq:delta}
\end {equation}
where $\{ t^A \}$ are $\Nc \times \Nc$
basis matrices for the Lie algebra of $U(\Nc)$,
normalized to satisfy
$
    \tr \> t^A t^B = \half \, \delta^{AB}
$
and
\begin{equation}
    \sum_{A=1}^{\Nc^2} \,
    (t^A)_{ij} \, (t^A)_{kl} = \half \, \delta_{il} \, \delta_{jk}
\label{eq:generator sum}
\end{equation}

\noindent
Because link variables are unitary, 
\begin {equation}
    \delta^A_\l \> u[\bar\l'] = -\delta_{\l\l'} \, u[\bar \l]\> t^A  \,.
\label {eq:delta2}
\end {equation}
Invariance of the Haar measure implies that the integral of any
variation vanishes,
\begin {equation}
    \int d\mu_0 \; \delta^A_\l \> ({\rm anything}) = 0 \,.
\end {equation}
Choosing `anything' to be $e^{S} \> \delta^A_\l \, (W[C])$,
and summing over the Lie algebra index $A$
(which will not be indicated explicitly),
gives the identity
\begin{equation}
    \Bigl\langle \delta_\l^A \delta_\l^A \, W[C] \Bigr\rangle 
    + \Bigl\langle \Bigl(\delta_\l^A \, W[C] \Bigr) 
    {\sum_{p \in \Lambda}}' \,
    \half \beta_p \,
    \Bigl(\delta^A_\l \> \tr \,(u[\partial p]{+}u[\partial p]^\dagger)\Bigr)
    \Bigr\rangle = 0 \ .
\label{eq:one link LE}
\end{equation}
This identity is easiest to visualize in cases where
the loop $C$ traverses the link $\l$ only once.
For such cases, the identity (\ref{eq:generator sum})
implies that the first term is just $\frac 12 \Nc$ times
the expectation value of $W[C]$.
The second term of the identity (\ref {eq:one link LE}) generates
terms in which plaquettes which also traverse the link $\l$
are `stitched' into the loop $C$ in all possible ways.
If both the loop $C$ and some plaquette boundary $\partial p$
contain the (directed) link $\l$, then
\begin {equation}
    \Bigl\langle
	\Bigl(\delta^A_\l \, W[C] \Bigr)
	\Bigl( \delta^A_\l \, \tr\,(u[\partial p] + u[\partial p]^\dagger)\Bigr)
    \Bigr\rangle
    =
    \half
    \left[
    \langle W[(\partial p) C] \rangle
    -
    \langle W[(\overline {\partial p}) C] \rangle
    \right] .
\label {eq:stitch}
\end {equation}
For the concatenation of $(\partial p)$ and $C$ to make sense,
both loops $C$ and $(\partial p)$
are to be regarded as starting with link $\l$
[and hence $(\overline {\partial p})$ ends with link $\bar\l\,$].
An example is shown in Fig.~\ref{fig:loop eq}.

\begin{FIGURE}[t]
    {
    \setlength{\unitlength}{2600sp}%
    %
    \begin{picture}(9624,3284)(439,-6003)
    \thinlines
    {\put(901,-3061){\line( 1, 0){900}}
    \put(1801,-3061){\line( 0,-1){600}}
    \put(1801,-3661){\line( 1, 0){600}}
    \put(2401,-3661){\line( 0,-1){600}}
    \put(2401,-4261){\line(-1, 0){1500}}
    \put(901,-4261){\line( 0, 1){1200}}}%
    {\put(901,-3061){\vector( 0,-1){750}}}%
    {\put(901,-4261){\vector( 1, 0){825}}}%
    {\put(2401,-4261){\vector( 0, 1){450}}}%
    {\put(2401,-3661){\vector(-1, 0){450}}}%
    {\put(1801,-3661){\vector( 0, 1){375}}}%
    {\put(1801,-3061){\vector(-1, 0){525}}}%
    {\put(701,-3061){\line(-1,-4){150}}
    \put(551,-3661){\line( 1,-4){150}}}%
    {\put(2601,-3061){\line( 1,-4){150}}
    \put(2751,-3661){\line(-1,-4){150}}}%
    {\put(4801,-3061){\line( 0,-1){1200}}
    \put(4801,-4261){\line( 1, 0){1500}}
    \put(6301,-4261){\line( 0, 1){600}}
    \put(6301,-3661){\line(-1, 0){600}}
    \put(5701,-3661){\line( 0, 1){600}}
    \put(5701,-3061){\line(-1, 0){300}}
    \put(5401,-3061){\line( 0, 1){300}}
    \put(5401,-2761){\line(-1, 0){300}}
    \put(5101,-2761){\line( 0,-1){300}}
    \put(5101,-3061){\line(-1, 0){300}}}%
    {\put(4801,-3061){\vector( 0,-1){750}}}%
    {\put(6301,-3661){\vector(-1, 0){450}}}%
    {\put(6301,-4261){\vector( 0, 1){450}}}%
    {\put(4801,-4261){\vector( 1, 0){825}}}%
    {\put(5701,-3661){\vector( 0, 1){375}}}%
    {\put(5701,-3061){\vector(-1, 0){225}}}%
    {\put(5401,-3061){\vector( 0, 1){225}}}%
    {\put(5401,-2761){\vector(-1, 0){225}}}%
    {\put(5101,-2761){\vector( 0,-1){225}}}%
    {\put(5101,-3061){\vector(-1, 0){200}}}%
    {\put(4601,-3061){\line(-1,-4){150}}
    \put(4451,-3661){\line( 1,-4){150}}}%
    {\put(6501,-3061){\line( 1,-4){150}}
    \put(6651,-3661){\line(-1,-4){150}}}%
    {\put(8101,-3061){\line( 0,-1){1200}}
    \put(8101,-4261){\line( 1, 0){1500}}
    \put(9601,-4261){\line( 0, 1){600}}
    \put(9601,-3661){\line(-1, 0){600}}
    \put(9001,-3661){\line( 0, 1){600}}
    \put(9001,-3061){\line(-1, 0){225}}
    \put(8776,-3061){\line( 0, 1){ 75}}
    \put(8776,-2986){\line(-1, 0){375}}
    \put(8401,-2986){\line( 0, 1){225}}
    \put(8401,-2761){\line( 1, 0){300}}
    \put(8701,-2761){\line( 0,-1){300}}
    \put(8701,-3061){\line(-1, 0){600}}}%
    {\put(8101,-3061){\vector( 0,-1){750}}}%
    {\put(8176,-4261){\vector( 1, 0){825}}}%
    {\put(9601,-4261){\vector( 0, 1){450}}}%
    {\put(9601,-3661){\vector(-1, 0){450}}}%
    {\put(9001,-3661){\vector( 0, 1){375}}}%
    {\put(9001,-3061){\vector(-1, 0){175}}}%
    {\put(8626,-2986){\vector(-1, 0){150}}}%
    {\put(8401,-2986){\vector( 0, 1){175}}}%
    {\put(8401,-2761){\vector( 1, 0){225}}}%
    {\put(8701,-2761){\vector( 0,-1){175}}}%
    {\put(8701,-3061){\vector(-1, 0){525}}}%
    {\put(7883,-3057){\line(-1,-4){150}}
    \put(7733,-3657){\line( 1,-4){150}}}%
    {\put(9801,-3061){\line( 1,-4){150}}
    \put(9951,-3661){\line(-1,-4){150}}}%
    {\put(4801,-4861){\line( 0,-1){1200}}
    \put(4801,-6061){\line( 1, 0){1500}}
    \put(6301,-6061){\line( 0, 1){600}}
    \put(6301,-5461){\line(-1, 0){600}}
    \put(5701,-5461){\line( 0, 1){600}}
    \put(5701,-4861){\line(-1, 0){300}}
    \put(5401,-4861){\line( 0,-1){300}}
    \put(5401,-5161){\line(-1, 0){300}}
    \put(5101,-5161){\line( 0, 1){300}}
    \put(5101,-4861){\line(-1, 0){300}}}%
    {\put(4801,-4861){\vector( 0,-1){750}}}%
    {\put(4801,-6061){\vector( 1, 0){825}}}%
    {\put(6301,-6061){\vector( 0, 1){450}}}%
    {\put(6301,-5461){\vector(-1, 0){450}}}%
    {\put(5701,-5461){\vector( 0, 1){375}}}%
    {\put(5701,-4861){\vector(-1, 0){225}}}%
    {\put(5401,-4861){\vector( 0,-1){225}}}%
    {\put(5326,-5161){\vector(-1, 0){175}}}%
    {\put(5101,-5161){\vector( 0, 1){225}}}%
    {\put(5026,-4861){\vector(-1, 0){175}}}%
    {\put(4583,-4857){\line(-1,-4){150}}
    \put(4433,-5457){\line( 1,-4){150}}}%
    {\put(6501,-4861){\line( 1,-4){150}}
    \put(6651,-5461){\line(-1,-4){150}}}%
    {\put(8101,-4861){\line( 0,-1){1200}}
    \put(8101,-6061){\line( 1, 0){1500}}
    \put(9601,-6061){\line( 0, 1){600}}
    \put(9601,-5461){\line(-1, 0){600}}
    \put(9001,-5461){\line( 0, 1){600}}
    \put(9001,-4861){\line(-1, 0){225}}
    \put(8776,-4861){\line( 0,-1){ 75}}
    \put(8776,-4936){\line(-1, 0){375}}
    \put(8401,-4936){\line( 0,-1){225}}
    \put(8401,-5161){\line( 1, 0){300}}
    \put(8701,-5161){\line( 0, 1){300}}
    \put(8701,-4861){\line(-1, 0){600}}}%
    {\put(8101,-4861){\vector( 0,-1){750}}}%
    {\put(8176,-6061){\vector( 1, 0){825}}}%
    {\put(9601,-6061){\vector( 0, 1){450}}}%
    {\put(9601,-5461){\vector(-1, 0){450}}}%
    {\put(9001,-5461){\vector( 0, 1){375}}}%
    {\put(9001,-4861){\vector(-1, 0){175}}}%
    {\put(8626,-4936){\vector(-1, 0){150}}}%
    {\put(8401,-4936){\vector( 0,-1){175}}}%
    {\put(8401,-5161){\vector( 1, 0){225}}}%
    {\put(8701,-5161){\vector( 0, 1){175}}}%
    {\put(8701,-4861){\vector(-1, 0){525}}}%
    {\put(7901,-4861){\line(-1,-4){150}}
    \put(7751,-5461){\line( 1,-4){150}}}%
    {\put(9801,-4861){\line( 1,-4){150}}
    \put(9951,-5461){\line(-1,-4){150}}}%
    \put(3050,-3730){\LARGE{$= \frac{\beta_p}{2\Nc} \!
			\left\{ \vphantom {\vbox to 1cm {\vfil}}
			\right.$}}
    \put(10051,-5530){\LARGE{$\left.
			\vphantom {\vbox to 1cm {\vfil}} \right\}$}}
    \put(3900,-5530){\LARGE{$+$}}
    \put(7050,-3730){\LARGE{$-$}}
    \put(7050,-5530){\LARGE{$-$}}
    \end{picture}
    \caption{
    The loop equation for a non-self-intersecting loop
    in pure gauge theory,
    when only one link (in the middle of the top edge) is varied.
    In this example, the lattice is
    two-dimensional and the coupling $\beta_p$
    is the same for all plaquettes.
    To aid visualization, here and in the following figures,
    links which are multiply traversed are shown slightly offset.
    Arrows on the loop indicate the direction of traversal,
    not the orientation of lattice links.
    On the right-hand side of the equation,
    ``untwisted'' plaquettes are attached to the 
    loop with a plus sign, and ``twisted'' plaquettes
    are attached with a minus sign.
    }
    \label {fig:loop eq}
    }
\end{FIGURE}

If the loop $C$ traverses link $\l$ more than once then
there are additional contributions generated by the first term in
(\ref {eq:one link LE}) in which the loop is cut apart into two
separate sub-loops.
Loops which multiply traverse some link
(in either direction) will be referred to as ``self-intersecting''.%
\footnote
    {%
    Hence, a loop may pass through a given site more than once and still
    be non-self-intersecting, provided it does not multiply traverse any link.
    }
As an example, if $C = \l C' \bar\l C''$, where $C'$ and $C''$ are
closed loops which do not contain link $\l$, then
\begin {equation}
    \frac 1\Nc \,
    \Bigl\langle \delta^A_\l \delta^A_\l \, W[C] \Bigr\rangle
    =
    \langle W[C] \rangle - \langle W[C'] \, W[C''] \rangle \,.
\end {equation}
Similarly, if $C = C' C''$ where $C'$ and $C''$
are non-self-intersecting closed loops both of which start with link $\l$,
then
\begin {equation}
    \frac 1\Nc \,
    \Bigl\langle \delta^A_\l \delta^A_\l \, W[C] \Bigr\rangle
    =
    \langle W[C] \rangle + \langle W[C'] \, W[C''] \rangle \,.
\end {equation}
These cases are illustrated in Figure \ref {fig:split1}.

\begin{FIGURE}
    {
    \hbox to \textwidth {\hfil
    \psfrag{l1}{\large{$\!\!\displaystyle \frac 1\Nc \,
		\delta_\l^A \delta_\l^A \qquad\qquad\qquad\, =$}}
    \psfrag{l2}{\small$\l$}
    \includegraphics[width=0.7\textwidth]{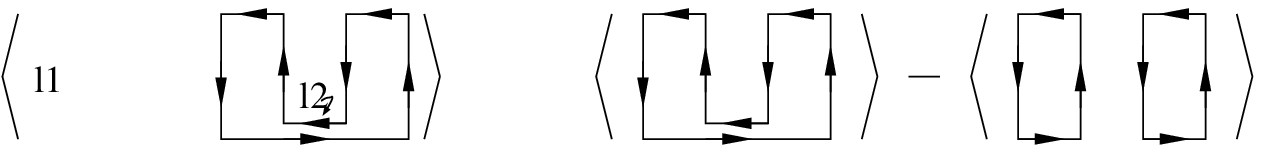}
    \hfil}
    \\[10pt]
    \hbox to \textwidth {\hfil
    \psfrag{l1}{\large{$\!\!\displaystyle \frac 1\Nc \,
    		\delta_\l^A \delta_\l^A \qquad\qquad\qquad\, =$}}
    \psfrag{l2}{\small$\l$}
    \includegraphics[width=0.7\textwidth]{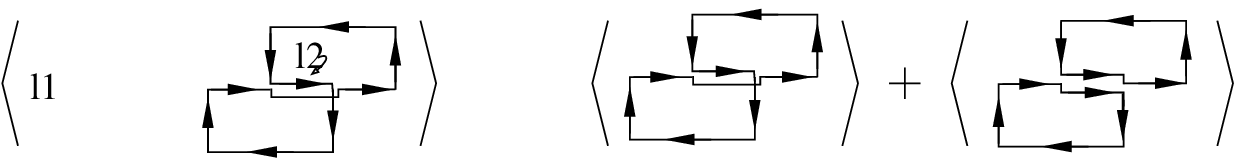}
    \hfil}
    \caption
	{
	Second variations of $W[C]$, when the link $\l$ being varied
	is traversed twice by the loop $C$ in opposite directions
	(above) or in the same direction (below).
	}
    \label {fig:split1}
    }
\end{FIGURE}

The identity (\ref {eq:one link LE}) depends, by construction,
on both the loop $C$ and the choice of link which is varied.
It will be simpler and more convenient
to instead sum over the varied link $\l$.
This will produce a single identity for each loop $C$,
which we will call `the loop equation for $W[C]$'.
As discussed below,
the resulting minimal set of loop equations will be sufficient to
determine the expectation values of all Wilson loops in the strong
coupling phase of the theory.
Summing over all links in the original identity (\ref{eq:one link LE})
yields the loop equation for the Wilson loop $W[C]$,
\begin {equation}
    0 =
    \Bigl\langle
	{\sum_{\l \in \Lambda}}' 
	\delta^A_\l \delta^A_\l \, W[C]
    \Bigr\rangle
    +
    \Bigl\langle
	{\sum_{\l \in \Lambda}}'
	\Bigl(\delta^A_\l \, W[C] \Bigr)
	{\sum_{p \in \Lambda}}' \,
	\half \beta_p \,
	\Bigl(\delta^A_\l \> \tr \,(u[\partial p]{+}u[\partial p]^\dagger)\Bigr)
    \Bigr\rangle
    \,.
\label {eq:loop eq 1}
\end {equation}
Non-zero contributions arise only for links $\l$ which are
traversed (in either direction) by the loop $C$ and, in the second term,
only for plaquettes whose boundaries also traverse the link $\l$.
The first term (divided by $\Nc$), now yields
\begin {equation}
    \frac 1\Nc \,
    \Bigl\langle
	{\sum_{\l \in \Lambda}}' \> \delta^A_\l \delta^A_\l \, W[C]
    \Bigr\rangle
    =
    \half |C| \, \langle W[C] \rangle \; {}
    + 
    \kern -5pt \sum_{\rm self-intersections} \kern-5pt
    \pm
    \langle W[C'] \, W[C''] \rangle \,,
\end {equation}
where $|C|$ denotes the length (total number of links) of $C$, and
$C'$ and $C''$ are the two sub-loops produced by reconnecting
$C$ at a given self-intersection.%
\footnote
    {
    If loop $C$ traverses a link $\l$ (in either direction) $K$ times,
    then the sum over self-intersections includes $K(K{-}1)/2$ terms
    generated by reconnecting each distinct pair of traversals of this link.
    }
Here, and subsequently, the upper sign applies if the two traversals
of the link at the intersection are in the same direction,
and the lower sign if the traversals are in opposite directions.
The second expectation in the identity (\ref {eq:loop eq 1}) generates
terms in which plaquettes which share any link with the loop $C$
are `stitched' into $C$ in all possible ways.
Combining these terms, the identity (\ref {eq:loop eq 1}),
divided by $\Nc$, may be re-written as%
\begin {eqnarray}
    \half |C| \, \langle W[C] \rangle
    &=&
    \sum_{\l \subset C} \>
    \sum_{p | \l \subset \partial p}
    \frac {\beta_p}{4\Nc}
    \left[
    \langle W[(\overline {\partial p}) C] \rangle
    -
    \langle W[(\partial p) C] \rangle
    \right]
\nonumber \\ &+&
    \kern -5pt \sum_{\rm self-intersections} \kern-5pt
    \mp
    \langle W[C'] \, W[C''] \rangle \,.
\label {eq:loop eq 2}
\end {eqnarray}
The sum over links $\l \subset C$ includes
all links contained in $C$,
with each link oriented according to the direction in which
it is traversed.
If the loop multiply traverses some link,
then the sum includes a separate term for each traversal.
The plaquette sum includes plaquettes oriented such
that their boundary contains link $\l$, not $\bar\l$.
The meaning of the first term on the
right-hand side of (\ref{eq:loop eq 2}) is
again simple: for every link of the loop,
``untwisted'' plaquettes (in all possible directions) 
are attached with a plus sign, 
and ``twisted'' plaquettes with a minus sign,
just like in Fig.~\ref{fig:loop eq}.

For any $U(\Nc)$ lattice gauge theory,
the loop equation (\ref {eq:loop eq 2}) is an exact identity relating
the expectation value of any Wilson loop to the expectation values of
modified loops with inserted plaquettes, and expectation values
of products of loops generated by reconnecting the original loop
at any self-intersections.

\subsection [Large $\Nc$ limit, strong-coupling expansion, and uniqueness]
    {Large $\bm \Nc$ limit, strong-coupling expansion, and uniqueness}

If the plaquette weights $\beta_p$ are scaled with the number of colors
so that $\betatilde_p \equiv \beta_p / \Nc$ is held fixed%
\footnote
    {%
    This is the same as the usual 't Hooft scaling in which $g^2 \Nc$
    (with $g$ the continuum gauge coupling) is held fixed,
    since the lattice coupling $\beta_p \sim 1/g^2$.
    }
as $\Nc \to \infty$,
then the loop equation (\ref {eq:loop eq 2}) becomes purely geometric
with no explicit $\Nc$ dependence.
It is known that the resulting large-$\Nc$ limit is a type of classical limit
\cite {LGY-largeN, Witten-largeN}
in which expectation values of products of Wilson loops factorize
into single-loop expectations, up to corrections subleading in $1/\Nc$,
\begin {equation}
    \langle W[C'] \, W[C''] \rangle
    =
    \langle W[C'] \rangle \, \langle W[C''] \rangle + O(1/\Nc^2) \,.
\end {equation}
Consequently, to leading order in the large $\Nc$ limit,
the loop equations (\ref {eq:loop eq 2}) become a set of closed,
non-linear equations only involving expectation values of single loops
\cite {MM},
\begin {eqnarray}
    \half |C| \, \langle W[C] \rangle
    &=&
    \sum_{\l \subset C} \>
    \sum_{p | \l \subset \partial p}
    \coeff 14 \, \betatilde_p
    \left[
    \langle W[(\overline {\partial p}) C] \rangle
    -
    \langle W[(\partial p) C] \rangle
    \right]
\nonumber \\ &+&
    \kern -5pt \sum_{\rm self-intersections} \kern -5pt
    \mp
    \langle W[C'] \rangle \, \langle W[C''] \rangle + O(1/\Nc^2) \,.
\label {eq:loop eq 3}
\end {eqnarray}
This closed set of equations
completely determines the expectation values of Wilson loops
in the large-$\Nc$ limit,
at least in the phase of the theory which is continuously connected
to strong coupling (small $\betatilde_p$).
To prove this rigorously, it is sufficient to
note that simply iterating the loop equations (\ref {eq:loop eq 3})%
\footnote
    {%
    Starting with $\langle W[C] \rangle = 0$ for all loops,
    except the trivial zero-length loop which is unity,
    $\frac 1\Nc \, \tr \, {\bm 1} = 1$.
    }
generates the lattice strong-coupling expansion ---
the expansion of expectation values in powers of the
plaquette weights $\betatilde_p$.
The significance of this follows from the fact that the
strong coupling expansion (unlike weak coupling perturbation theory)
is known to have a non-zero radius of convergence
\cite {OS, Seiler, LGY-lattice}.
Combined with the uniqueness of analytic continuation,
this shows that the loop equations (\ref {eq:loop eq 3})
uniquely determine the large-$\Nc$ expectation values of Wilson loops
throughout the strong coupling phase of the theory.

In the large-$\Nc$ limit, all Wilson-action $U(\Nc)$ lattice gauge theories
(even in finite volume) are believed to possess a third-order phase transition
which is an artifact of the $\Nc\to\infty$ limit
\cite {GW,Friedan,Neuberger1,Neuberger2}.
This phase transition is driven by the behavior of the distribution
of eigenvalues of elementary plaquettes $u[\partial p]$.
For sufficiently strong coupling (small $\betatilde$) the eigenvalue
distribution is non-zero on the entire unit circle,
while for sufficiently weak coupling (large $\betatilde$) the support
of this distribution lies only on a subset of the circle.
On the weak-coupling side of this phase transition,
it is sufficient
(at least in simple models involving one, two, or three plaquettes
\cite {Friedan})
to supplement the loop equations (\ref {eq:loop eq 3}) with the
trivial inequalities
\begin {equation}
    |\langle W[C] \rangle| \le 1 \,,
\label {eq:inequalities}
\end {equation}
in order to select the correct root of the loop equations.

\subsection {Multi-loop connected correlators}

The preceding formulation of loop equations may be easily extended
to connected correlators involving a product of two or more Wilson loops.
This extension will be needed for our later discussion,
since the particle spectrum of a theory can only be extracted
from two-loop correlators, not from single loop expectation values.

For notational convenience, we will define rescaled $k$-loop connected
correlators,
\begin {equation}
    \dlangle W[C_1] W[C_2] \cdots W[C_k] \drangle
    \equiv
    \Nc^{2(k-1)} \, \langle W[C_1] W[C_2] \cdots W[C_k] \rangle_{\rm conn.}
\label {eq:multi-loop}
\end {equation}
The connected part of $k$-loop correlators vanish as $O(1/\Nc^{2(k-1)})$
relative to the totally disconnected part \cite {Witten-largeN, Coleman-largeN}.
Consequently,
the overall factors of $\Nc$ inserted in (\ref {eq:multi-loop}) allow
the rescaled connected correlators $\dlangle W[C_1] \cdots W[C_k] \drangle$ to
have finite, non-trivial large $\Nc$ limits.

For two-loop correlators, the natural generalization of
the identity (\ref {eq:loop eq 1}) is
\begin {eqnarray}
    0 &=&
    \int d\mu_0 \> {\sum_{\l \in \Lambda}}' \> \delta^A_\l
    \left\{
	e^{S} \> \delta^A_\l
	\left[\strut
	    \Bigl(W[C_1] {-} \langle W[C_1] \rangle\Bigr) \,
	    \Bigl(W[C_2] {-} \langle W[C_2] \rangle\Bigr)
	\right]
    \right\}
\nonumber \\ &=&
    \Bigl\langle
	{\sum_{\l \in \Lambda}}'
	\Bigl[
	    \delta^A_\l \delta^A_\l \, W[C_1]
	    +
	    {\sum_{p \in \Lambda}}' \>
	    \half \Nc \beta_p
	    \Bigl(
		\delta^A_\l \, W[C_1]
	    \Bigr)
	    \delta^A_\l
	    \Bigl(W[\partial p] {+} W[\overline {\partial p}]\Bigr)
	\Bigr]
	\Bigl( W[C_2] {-} \langle W[C_2] \rangle \Bigr)
    \Bigr\rangle
\nonumber \\ && {} +
    \Bigl\langle
	{\sum_{\l \in \Lambda}}'
	\Bigl( \delta^A_\l \, W[C_1] \Bigr)
	\Bigl( \delta^A_\l \, W[C_2] \Bigr)
	\Bigr\rangle
    +
    (C_1 \leftrightarrow C_2) \,.
\label {eq:two loop}
\end {eqnarray}
The expectation value in the second line of the result is only non-zero if
loops $C_1$ and $C_2$ intersect
({\em i.e.}, both loops traverse a common link);
if so then this term is $1/\Nc$ times the expectation value of a single Wilson
loop produced by reconnecting loops $C_1$ and $C_2$ at their
mutual intersection(s). 
The disconnected part of the expectation in the first line of the result
vanishes identically (both factors are zero),
so what survives comes from connected two-loop correlators, as desired.
The resulting contribution is also $O(1/\Nc)$, due to the
overall factor of $\Nc$ (either explicit, or hidden in the action
of $\delta^A_\l \delta^A_\l$ on $W[C_1]$) multiplying an $O(1/\Nc^2)$
connected two-loop correlator.

Multiplying the identity (\ref {eq:two loop}) by an overall factor of $\Nc$ and
sending $\Nc\to\infty$ yields
\begin {eqnarray}
    \half (|C_1|{+}|C_2|) \, \dlangle W[C_1]\,W[C_2] \drangle
    &=&
    \!\sum_{\l \subset C_1} \, \sum_{p | \l \subset \partial p} \!
    \coeff 14 \, \betatilde_p
    \Bigl[
	\dlangle W[(\overline {\partial p})C_1] \, W[C_2] \drangle
	{-}
	\dlangle W[(\partial p)C_1] \, W[C_2] \drangle
    \Bigr]
\nonumber \\[2pt] &+&
    \kern -5pt
    \sum_{{\scriptstyle {\rm self-intersections} \atop \scriptstyle (C_1)}}
    \kern -5pt
    \mp \Bigl[
	    \dlangle W[C_1'] W[C_2] \drangle \,
	    \langle W[C_1''] \rangle
	    + (C_1' \leftrightarrow C_1'')
	\Bigr]
\nonumber \\ &+&
    \kern -5pt \sum_{\scriptstyle {\rm mutual-intersections}} \kern -5pt
	\mp \half \, \langle W[C_1 C_2] \rangle + O(1/\Nc^2)
\nonumber \\[5pt] &+&
    (C_1 \leftrightarrow C_2) \,.
\label {eq:2-loop eq}
\end {eqnarray}
Once again, in each plaquette insertion term both the loop $C_1$
and the plaquette boundary $\partial p$ are to be regarded as 
starting with link $\l$.
In the mutual intersection terms,
$C_1$ and $C_2$
are to be regarded as either starting with the intersection link
or ending with its conjugate;
the upper (lower) sign applies if both loops traverse the intersection link
in the same (opposite) direction.
The omitted $O(1/\Nc^2)$ piece involves the
fully-connected three-loop correlator
$\langle W[C_1'] W[C_1''] W[C_2] \rangle_{\rm conn}$.
This may be dropped in the large-$\Nc$ limit 
because connected three-loop correlators vanish faster (by $1/\Nc^2$)
than two-loop correlators.

The result (\ref {eq:2-loop eq}) is a set of inhomogeneous
linear equations for connected two-loop correlators.
Just like the loop equations for single Wilson loops,
these connected correlator loop equations may be solved
iteratively (starting with all two-loop correlators equal to zero)
to produce a strong coupling expansion with a non-zero radius of convergence.
Hence, this set of equations (together with the single loop equations)
completely determine two-loop connected correlators,
at least in the phase of the theory which is continuously connected
to strong coupling [and presumably beyond as well, when supplemented
with the inequalities (\ref {eq:inequalities})].

The extension to higher multi-loop correlators is completely analogous,
and will not be discussed explicitly.

\section {Adjoint matter fields}
\label{sec:Adjoint matter}

\subsection {Lattice discretization}

At every site $s$ of the lattice,
we now add $\Ns$ independent scalar variables
$\{\phi_a[s]\}$ ($a = 1,\cdots,\Ns$),
and $\Nf$ pairs of fermionic variables
$\{ \psi_b[s], \bar\psi_b[s]\}$ ($b = 1,\cdots,\Nf$),
all transforming in the adjoint representation of the $U(\Nc)$ gauge group.
The scalars $\phi_a[s]$ are
complex $\Nc \times \Nc$ matrices,%
\footnote
    {
    Since the adjoint representation is a real representation,
    one could introduce scalar variables as $\Nc \times \Nc$ Hermitian
    matrices.
    We choose to use complex scalars instead,
    so that the resulting theory will have a $U(\Ns)$ global symmetry.
    This will allow us to apply non-trivial orbifold projections
    (even when $\Ns = 1$),
    as discussed in the next section.
    }
while the fermions $\psi_b[s]$ and $\bar\psi_b[s]$ are 
$\Nc \times \Nc$ matrices of independent Grassmann variables.
Although we were more general in the last section,
henceforth we will assume that the lattice $\Lambda$ is a
simple cubic lattice.

The measure for the theory has the usual form (\ref {eq:dmu}),
where the decoupled measure $d\mu_0$ is now the product of Haar
measure for every link variable and independent flat measures
for all the scalar and Grassmann variables,%
\footnote
    {
    $d\phi_a[s] \> d\phi_a^\dagger[s]$ should be understood
    as denoting independent integration over each of the $2\Nc^2$
    real degrees of freedom contained in $\phi[s]$.
    Similarly, $d\psi_b[s]$ and $d\bar\psi_b[s]$ should be
    understood
    as denoting independent integration over each of the $\Nc^2$
    Grassmann degrees of freedom contained in $\psi_b[s]$
    and $\bar\psi_b[s]$, respectively.
    }
\begin {equation}
    d\mu_0 =
    \Bigl( {\prod_{\l\in\Lambda}}' \> du[\l] \Bigr)
    \Bigl( \prod_{s\in\Lambda} \>\prod_{a=1}^{\Ns}
	d\phi_a[s] \> d\phi_a^\dagger[s]
    \Bigr)
    \Bigl( \prod_{s\in\Lambda} \>\prod_{b=1}^{\Nf}
	    d\psi_b[s]\>d\bar\psi_b[s]
    \Bigr) \,.
\end {equation}

The action is the sum of the pure-gauge Wilson action (\ref {eq:S-YM}),
which we now denote as $S_{\rm gauge}$, plus matter field contributions,
\begin {equation}
    S = S_{\rm gauge} + S_{\rm scalar} + S_{\rm fermion} \,.
\label {eq:Smatter}
\end {equation}
The scalar action will have the natural nearest-neighbor coupling plus
some local potential energy,
\begin {eqnarray}
    \!\!\!\!
    S_{\rm scalar}
    &=&
    \Nc^2 \, \Biggl\{
    \frac \kappa 2
    \sum_{\l=\langle ss' \rangle \in\Lambda}\kern-5pt
	\tr \left(
	\phi_a^{\smash{\dagger}}[s] \, u[\l] \,
	\phi_a[s'] \, u[\bar \l]
	\right)
	\!/\Nc
    -
    \sum_{s\in\Lambda} \>
    V\!\left[
	\vphantom {\bar\phi}
	\tr \left(\phi_a^{\smash{\dagger}}[s]\,\phi_a[s]\right)
	\!/\Nc
    \right]
    \Biggr\} \,
    .
\label {eq:Sscalar}
\end {eqnarray}
The notation $\l=\langle ss' \rangle$ means that $\l$ is the link
which runs from site $s$ to neighboring site $s'$;
this sum runs over both orientations of every link.
There is an implied sum over the repeated ``flavor'' index $a$.
We have chosen the scalar action to have a $U(\Ns)$ global symmetry;
the specific form of the scalar potential could be generalized
at the cost of extra notational complexity.
To ensure integrability of the full measure $d\mu$, the potential
$V[\chi]$ should rise unboundedly for large arguments.
For later convenience, we have inserted factors of $\Nc$ so that both
the ``hopping parameter'' $\kappa$
and the functional form of $V[\chi]$
may be kept fixed as $\Nc \to \infty$
(with $\Ns$ fixed).

The fermion action is
\begin {eqnarray}
    S_{\rm fermion}
    &=&
    \Nc \, \Biggl\{
    \frac \kappa {2i}
    \sum_{\l=\langle ss' \rangle \in\Lambda}\kern-5pt
	\tr 
	\left(
	    \bar\psi_b[s] \, \eta[\l] \, u[\l] \, \psi_b[s'] \,
	    u[\bar \l]
	\right)
    -
    m \sum_{s\in\Lambda} \,
    \tr \left(\bar\psi_b[s] \, \psi_b[s]\right)
    \Biggr\}\,
    ,
\label {eq:Sfermion}
\end {eqnarray}
with an implied sum on the flavor index $b$.
We have chosen both scalars and fermions to have a common hopping parameter
$\kappa$.
This may always be arranged by suitably rescaling the scalar (or fermion)
variables.
Similarly, as long as the common fermion bare mass $m$ is non-zero,
it may be set to unity by an appropriate rescaling of variables.
Physical quantities only depend on the ratio $\kappa/m$;
hence large mass $m$ is equivalent to small hopping parameter $\kappa$. 
In the fermion action (\ref {eq:Sfermion}),
$\eta[\l]$ is an imaginary phase factor
assigned to each link in such a way
that the product of these phases around every plaquette is minus one,
$
    \eta[\partial p] = -1
$.
We will refer to it as the ``fermion flavor connection'';
as with any unitary connection, $\eta[\bar \l] \equiv \eta[\l]^\dagger$.
A specific realization is
\begin {equation}
    \eta[\l] = i \prod_{\nu < \mu} (-1)^{x_\nu} \,,
\end {equation}
if $\l$ is the link running in the $\hat e_\mu$ direction starting
from the site with coordinates ${x_\mu}$.

The choice (\ref {eq:Sfermion}) for discretizing fermion fields is known as
``staggered lattice fermions''
\cite {Susskind, Staggered};
it has the virtue of being notationally compact and not
cluttering expressions with extraneous Dirac indices and gamma matrices.%
\footnote
    {
    In $d$ spacetime dimensions, the naive discretization
    of a single Dirac fermion is equivalent to $2^{\lfloor d/2 \rfloor}$
    species of staggered fermions \cite{Staggered}
    (with ${\lfloor d/2 \rfloor}$ denoting the integer part of $d/2$).
    }
We have chosen the fermion action to have a $U(\Nf)$ global symmetry;
this assumption could be relaxed and
the bare mass $m$ replaced by an arbitrary mass matrix
at the cost of extra notational complexity.
We have again inserted factors of $\Nc$ in a manner which will
prove convenient when taking the large $\Nc$ limit
(with $\Nf$ fixed).


\subsection {Geometric encoding of observables}
\label {sec:encoding}

For theories with adjoint representation matter fields, the natural
gauge invariant observables are Wilson loops ``decorated'' with
arbitrary insertions of matter fields at sites through which the loop passes.
To formulate appropriate generalizations of loop equations,
a required first step is adopting some scheme for unambiguously labeling
arbitrarily decorated loops.

Consider, for the moment, a theory with only fermionic matter fields.
One possibility would be to define
\begin {equation}
    W[\Gamma_1, \Gamma_2, \cdots, \Gamma_K]_{b_1 \bar b_2\cdots b_K}
    =
    \frac 1\Nc \> \tr \left(
	\psi_{b_1}[s_1] \, u[\Gamma_1] \,
	\bar\psi_{b_2}[s_2] \, u[\Gamma_2] \cdots u[\Gamma_{K-1}] \,
	\psi_{b_K}[s_K] \, u[\Gamma_K]
	^{\vphantom{\dagger}}
    \right) ,
\label {eq:W-decorated}
\end {equation}
{\em etc}.
The $\Gamma_i$ are (in general) arbitrary open paths on the lattice which,
when concatenated, form a closed loop,
with $s_i$ the site at which segment $\Gamma_i$ begins.
Employing this notation is possible, but
(a) its excessively lengthy,
(b) it does not uniquely label observables (due to trace cyclicity), and
(c) one can do better.

A more concise, geometric labeling of observables may be
formulated if one considers a $(d{+}\Nf)$ dimensional lattice
constructed by tensoring the original lattice $\Lambda$ with
$\Nf$ copies of the integers,
$\Lambda' \equiv \Lambda \times \Z^{\Nf}$.
Each $d$-dimensional `slice' of $\Lambda'$ looks just like
$\Lambda$, except that $\Nf$ additional (oriented) links,
pointing into the $\Nf$ new dimensions,
now emanate from every site.
The basic idea is to treat the fermion variables $\{\bar\psi_b[s]\}$ as the
connection associated with links pointing into the new dimensions,
and $\{\psi_b[s]\}$ as the conjugate connection associated with the
oppositely directed links.
The connection on links pointing in directions lying in any of the original
$d$ dimensions is the initial gauge field $u[\l]$.
This is illustrated in Figure~\ref {fig:lattice 1}.
To write this more explicitly, let $\hat e_b$ denote unit vectors
pointing in each of the $\Nf$ new dimensions.
Links $\l' \in \Lambda'$ either point in a direction which lies in
the original $d$ dimensions,
in which case they may be labeled as $\l' = (\l,\vec n)$
[with $\l \in \Lambda$ and $\vec n \in \Z^{\Nf}$],
or they point in one of the new directions in which case
they may be labeled as $\l' = (s,\vec n,\pm\hat e_b)$
[with $s \in \Lambda$ and $\vec n \in \Z^{\Nf}$].
Let $\Z^{\Nf}_+$ denote the even sub-lattice of $\Z^{\Nf}$
(points whose coordinates sum to an even integer),
and $\Z^{\Nf}_-$ the odd sub-lattice.
We define a lattice link variable $v[\l']$ on $\Lambda'$
such that
\begin {equation}
    v[\l']
    =
    \left\{
	\begin {array}{ll}
	u[\l]
	\,, &\mbox { if }\l' = (\l,\vec n);
	\\[2pt]
	\bar \psi_b[s]
	    \,, &\mbox { if } \l' = (s,\vec n,+\hat e_b);
	\\[2pt]
	\psi_b[s]
	    \,, &\mbox { if } \l' = (s,\vec n,-\hat e_b).
	\end {array}
    \right.
\label {eq:v}
\end {equation}

\begin {FIGURE}[t]
    {
    \setlength{\unitlength}{4000sp}
    \begin{picture}(4524,1584)(1189,-3403)
    \thinlines
    {\put(3001,-3361){\line( 0, 1){600}}
    \put(3001,-2761){\line( 1, 0){600}}
    \put(3601,-2761){\line( 0,-1){600}}}
    %
    %
    {\put(3030,-3450){$s$}}
    {\put(3431,-3450){$s'$}}
    \put(2650,-3100){$\bar\psi[s]$}
    \put(3650,-3100){$\psi[s']$}
    \put(3190,-2680){$u[\l]$}
    {\put(3601,-3361){\line( 1, 0){600}}
    \put(4201,-3361){\line( 1, 1){1500}}
    \put(5701,-1861){\line(-1, 0){600}}
    \put(5101,-1861){\line(-1,-1){300}}
    }%
    {\put(4801,-2161){\line(-1, 0){2400}}
    \put(2401,-2161){\line(-1,-1){1200}}}%
    {\put(1201,-3361){\line( 1, 0){1800}}}%
    {\put(3001,-2761){\vector( 1, 0){430}}}%
    {\put(3601,-2761){\vector( 0,-1){375}}}%
    {\put(3601,-3361){\vector( 1, 0){375}}}%
    {\put(4201,-3361){\vector( 1, 1){975}}}%
    {\put(5701,-1861){\vector(-1, 0){450}}}%
    {\put(5101,-1861){\vector(-1,-1){225}}}%
    {\put(4801,-2161){\vector(-1, 0){1425}}}%
    {\put(2401,-2161){\vector(-1,-1){675}}}%
    {\put(1201,-3361){\vector( 1, 0){1200}}}%
    {\put(3001,-3361){\vector( 0, 1){375}}}%
    \put(5501,-3000){\vector( 0, 1){375}}
    \put(5501,-3000){\vector(-1,-1){250}}
    \put(5501,-3000){\vector( 1, 0){375}}
    \put(5550,-2800){{\small new dimension}}
    \put(5500,-3250){$\Lambda$}
    \end{picture}
    \caption
	{
	A closed loop in the extended lattice $\Lambda'$,
	for the case of one fermion flavor.
	As indicated, links pointing in the new dimension represent fermion
	variables $\bar\psi[s]$ and $\psi[s']$.
	The observable associated with this closed loop is
	$
	    \frac 1 \Nc \, \tr
	    \left(
		\bar\psi[s] \, u[\l] \, \psi[s'] \, u[\Gamma]
	    \right)
	$,
	with $\l$ the link running from site $s$ to site $s'$,
	and $\Gamma$ denoting the portion of the contour lying
	in the original lattice $\Lambda$ and running from site $s'$
	back to $s$.
	Only those decorated Wilson loops in which the number of
	$\psi$ and $\bar\psi$
	insertions coincide (for each flavor) form closed loops on $\Lambda'$.
	}
    \label {fig:lattice 1}
    }
\end {FIGURE}

\begin {FIGURE}[t]
    {
    \hbox to \textwidth {\hfil
    \setlength{\unitlength}{3947sp}
    \begin{picture}(3104,1104)(2989,-3403)
    \thinlines
    \put(3001,-3361){\line( 1, 0){600}}
    \put(3601,-3361){\line( 1, 1){450}}
    \put(4051,-2911){\line( 0, 1){600}}
    \put(4051,-2311){\line(-1, 0){ 75}}
    \put(3976,-2311){\line( 0,-1){600}}
    \put(3976,-2911){\line(-1, 0){525}}
    \put(3451,-2911){\line(-1,-1){450}}
    \put(3001,-3361){\vector( 1, 0){450}}
    \put(3601,-3361){\vector( 1, 1){300}}
    \put(4051,-2911){\vector( 0, 1){450}}
    \put(3976,-2311){\vector( 0,-1){525}}
    \put(3976,-2911){\vector(-1, 0){375}}
    \put(3451,-2911){\vector(-1,-1){300}}
    \put(4070,-3000){$s$}
    \put(4120,-2600){$\bar\psi[s]$}
    \put(3600,-2600){$\psi[s]$}
    \put(3550,-3150){$p$}
    \put(4550,-3150){$\neq$}
    \put(5500,-3150){$p$}
    \put(6070,-3000){$s$}
    \put(5001,-3361){\line( 1, 0){600}}
    \put(5601,-3361){\line( 1, 1){450}}
    \put(6050,-2911){\line(-1, 0){600}}
    \put(5451,-2911){\line(-1,-1){450}}
    \put(5001,-3361){\vector( 1, 0){400}}
    \put(5601,-3361){\vector( 1, 1){300}}
    \put(6050,-2911){\vector(-1, 0){400}}
    \put(5451,-2911){\vector(-1,-1){300}}
    \end{picture}
    \hfil
    }
    \caption
	{
	Backtracking ``stubs'' extending in the new directions do not cancel,
	since the associated ``link variables'' are non-unitary matter fields.
	The indicated loop on the left represents the observable
	$
	    \frac 1 \Nc \,
	    \tr \left(\bar\psi[s] \, \psi[s] \, u[\partial p]\right)
	$,
	where $p$ is a plaquette whose boundary passes through the site $s$.
	}
    \label {fig:backtracking}
    }
\end {FIGURE}

Now apply the standard definition of Wilson loops,
using the connection (\ref {eq:v}),
to arbitrary closed paths in the lattice $\Lambda'$.
More precisely, we define
\begin {equation}
    W[C] = \pm \, \tr \> v[C] \,,
\label {eq:W-extended}
\end {equation}
with the upper sign ($+$) applying if the path $C$ is written as a sequence
of links starting at a site in the even sub-lattice $\Z_+^{\Nf}$,
and the lower sign ($-$) if the path $C$ is written as a sequence
of links starting at a site in the odd sub-lattice $\Z_-^{\Nf}$.%
\footnote
    {
    If the overall $\pm$ sign were omitted then,
    due to the Grassmann nature of fermion variables,
    different choices for the starting site of a loop
    would correspond to observables with differing overall signs.
    (For observables with an even number of fermion insertions,
    moving a Grassmann variable from one end of the trace to the
    other requires an odd number
    of interchanges with other Grassmann variables.)
    With the definition (\ref {eq:W-extended}),
    the observable $W[C]$ depends only on the geometry of the loop
    $C \in \Lambda'$,
    not on the arbitrary choice of starting site.
    }
Each loop $C \in \Lambda'$ generates an observable resembling
the example (\ref {eq:W-decorated}) (or else a normal Wilson loop if $C$
lies entirely in a $d$-dimensional slice parallel to $\Lambda$),
up to an overall sign.
Note, however, that
\begin {equation}
    v[\l'] \, v[\bar\l'] \ne 1 \quad
    \hbox {if } \l' = (s,\vec n, \pm \hat e_b) \,,
\end {equation}
because the connection $v[\l']$ is not unitary
for links pointing in the $\Nf$ new directions.
Hence, backtracking ``stubs'' involving the new links do not cancel.
This is illustrated in Figure~\ref {fig:backtracking}.

Of course, a path $C$ in the extended lattice $\Lambda'$ which
includes $M$ links pointing in the $+\hat e_b$ direction must also
include $M$ links pointing in the $-\hat e_b$ direction
if it is to form a closed loop.
So this mapping of loops in the higher-dimensional lattice $\Lambda'$
onto observables of the form (\ref {eq:W-decorated}) only generates
observables which separately conserve the number of each staggered
fermion species.
This is adequate for some purposes, but it is insufficient if
one wishes to consider theories with non-diagonal or Majorana mass terms.
More importantly, it is inadequate even in theories where net fermion
number of each species is conserved, if one wishes to consider two (or higher)
point correlation functions in all possible flavor symmetry channels.

\begin{FIGURE}[t]
    {
    \hbox to \textwidth {\hfil
    \psfrag{s1}{$s$}
    \psfrag{s2}{$s'$}
    \psfrag{psi1}{$\bar\psi$}
    \psfrag{psi2}{$\ \bar\psi$}
    \psfrag{psi3}{$\bar\psi$}
    \psfrag{psi4}{$\ \psi$}
    \includegraphics[width=0.5\textwidth]{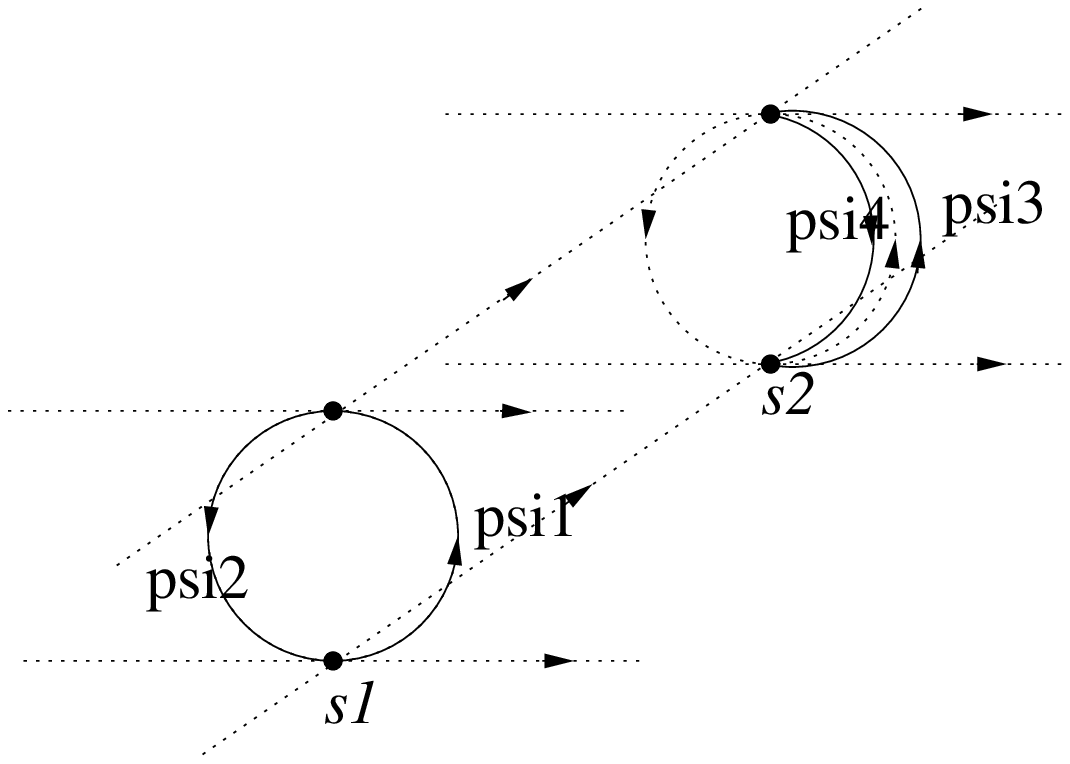}
    \hfil}
    \caption
	{
	Examples of closed loops on the
	minimally extended lattice $\bar\Lambda_{\rm f}$,
	for the case of one fermion flavor.
	Dotted lines represent positively oriented links.
	The observables associated with the indicated closed paths are
	$\frac 1 \Nc \, \tr \left(\bar\psi[s] \bar\psi[s] \right)$ and
	$\frac 1 \Nc \, \tr \left(\bar\psi[s'] \psi[s'] \right)$.
	Both are closed loops on $\bar\Lambda_{\rm f}$, 
	but only the second one would be a closed loop 
	on $\Lambda'$.
	All Wilson loops containing an even total number of
	fermion insertions form closed loops on $\bar\Lambda_{\rm f}$.
	}
    \label {fig:lattice 2}
    }
\end{FIGURE}

One may accommodate a larger class of observables by
appropriately identifying sites in $\Lambda'$, since this
enlarges the set of closed loops.
The smallest lattice, and the largest set of acceptable observables,
is produced by identifying all sites in $\Lambda'$ which differ by even
translations in $\Z^{\Nf}$
(those whose displacement vectors lie in $\Z^{\Nf}_+$).
The result is a lattice $\bar\Lambda_{\rm f} \equiv \Lambda' / \Z^{\Nf}_+$
whose sites are just $\Lambda \otimes Z_2$, but where each site is
connected to its $Z_2$ partner by $\Nf$ distinct, positively oriented links.
This is illustrated in Figure~\ref {fig:lattice 2}.
Let $\widetilde \Lambda$ denote the $Z_2$ image of the original
sublattice $\Lambda$, and let $\tilde s$ (or $\tilde \l$ or $\tilde p$) denote
the $Z_2$ partner of any site $s$ (or link $\l$ or plaquette $p$).
The reduction of the definition (\ref {eq:v}) of the lattice link variable is%
\footnote
    {
    $(s,+\hat e_b)$ denotes the positively oriented link which
    runs from site $s$ to $\tilde s$ in the direction $+\hat e_b$,
    and $(\tilde s,-\hat e_b)$ denotes the same link
    but in its opposite orientation.
    $(\tilde s,+\hat e_b)$ denotes the positively oriented link which
    runs from site $\tilde s$ to $s$ in the direction $+\hat e_b$,
    and $(s,-\hat e_b)$ denotes the same link in its opposite orientation.
    }
\begin {equation}
    v[\l']
    =
    \left\{
	\begin {array}{ll}
	u[\l] \,, &\hbox{either }\l' = \l \in \Lambda,
	\hbox{ or }\l' = \tilde \l \in \widetilde\Lambda;
	\\[2pt]
	\bar \psi_b[s] \,, &\hbox{either } \l'=(s,+\hat e_b), \; s \in \Lambda,
	\hbox{ or }
	\l'=(\tilde s,+\hat e_b), \; \tilde s \in \widetilde \Lambda;
	\\[2pt]
	\psi_b[s]\,,&\hbox{either }\l'=(s,-\hat e_b), \; s \in \Lambda,
	\hbox{ or }
	\l'=(\tilde s,-\hat e_b), \; \tilde s \in \widetilde \Lambda.
	\end {array}
    \right.
\label {eq:v2}
\end {equation}

All gauge invariant bosonic observables of the form (\ref {eq:W-decorated})
({\em i.e.}, those in which the total number of fermion insertions is even)
may now be represented by closed loops in $\bar\Lambda_{\rm f}$.%
\footnote
    {
    If $\widetilde C$ denotes the $Z_2$ image of a loop $C$
    (so that every link $\l$ in $C$ is replaced by its
    $Z_2$ partner $\tilde \l$),
    then the corresponding observables differ only by an overall sign,
    $W[\widetilde C] = -W[C]$.
    Hence, the set of all closed loops on $\bar\Lambda_{\rm f}$ represents
    all single-trace bosonic observables as well as their negations.
    \label {fn:Z2mirror}
    }
Gauge invariant fermionic observables do not correspond to closed loops
in the extended lattice $\bar\Lambda_{\rm f}$, but rather to open paths
whose endpoints are $Z_2$ images of each other ---
paths beginning at some site $s$ and ending at $\tilde s$.
As will be seen shortly, large $\Nc$ loop equations for either
expectation values or multi-loop correlators of bosonic observables
will not involve fermionic observables.
For simplicity, we will focus on the treatment of bosonic observables
in the following discussion, and relegate discussion of
correlators of fermionic observables to
Appendix~\ref{sec:Fermionic observables}.

``Gauge-fermion'' plaquettes
({\em i.e.}, plaquettes containing both fermion links and ordinary gauge links)
reproduce the hopping terms in the fermion action (\ref {eq:Sfermion}).
For the plaquette $p$ whose boundary contains the gauge link
$\bar\l = \langle s' s \rangle \in \Lambda$
and the fermion link $(s,+\hat e_b)$,
\begin {equation}
    \tr \left(v[\partial p]\right)
    =
    \tr \left(
    \bar\psi_b[s] \, u[\l] \, \psi_b[s'] \, u[\bar\l]
    \right) ,
\end {equation}
which coincides (up to the phase $\eta[\l]$)
with the hopping term in the action (\ref {eq:Sfermion}).

\begin {FIGURE}[t]
    {
    \hbox to \textwidth {\hfil
    \psfrag{s1}{$s$}
    \psfrag{s2}{$s'$}
    \psfrag{phi1}{$\phi^\dagger[s]$}
    \psfrag{phi2}{$\phi^\dagger[s]$}
    \psfrag{phi3}{$\phi^\dagger[s']$}
    \psfrag{phi4}{$\phi[s']$}
    \includegraphics[width=0.4\textwidth]{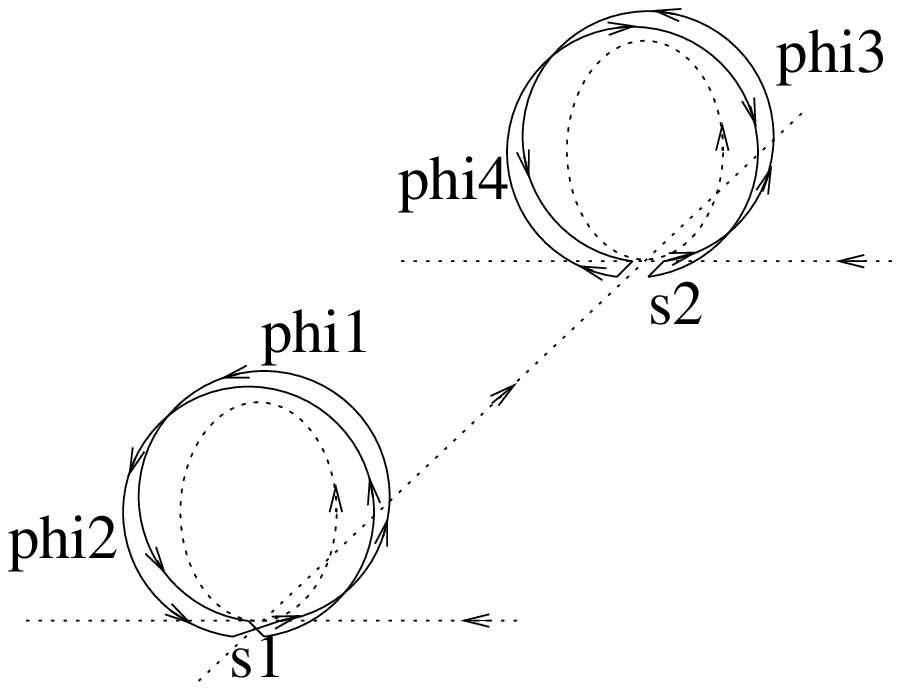}
    \hfil}
    \caption
	{
	Examples of closed loops on the
	extended lattice $\bar\Lambda_{\rm s}$ for the case
	of one scalar flavor.
	Dotted lines represent (oriented) links.
	The observables associated with the indicated closed paths are
	$\frac 1 \Nc \, \tr \left(\phi^\dagger[s] \phi^\dagger[s] \right)$ and
	$\frac 1 \Nc \, \tr \left(\phi^\dagger[s'] \phi[s'] \right)$.
	Wilson loops with any number of scalar insertions
	form closed loops on $\bar\Lambda_{\rm s}$.
	}
    \label {fig:lattice 3}
    }
\end {FIGURE}

The above strategy may be applied equally well
to theories with adjoint representation scalars.
With scalar insertions,
there are no subtleties concerning the overall sign of an observable.
Hence, one may divide the covering lattice $\Lambda'$ by the 
entire $\Z^{\Ns}$ translation group.
The net result is a lattice $\bar \Lambda_{\rm s}$ whose sites coincide
with those of $\Lambda$ but where every site now has $\Ns$
independent oriented links which connect the site to itself,
as illustrated in Figure~\ref {fig:lattice 3}.
The appropriate lattice link variable is now%
\footnote
    {
    $(s,+\hat e_a)$ denotes the ``scalar'' link which connects
    site $s$ to itself running in direction $\hat e_a$;
    $(s,-\hat e_a)$ is the opposite orientation of the same link.
    (If we had chosen our scalar variables to be Hermitian, then
    it would have been natural to regard these links as unoriented.)
    }
\begin {equation}
    v[\l']
    =
    \left\{
	\begin {array}{ll}
	u[\l] \,, &\l' = \l \in \Lambda;
	\\[2pt]
	\phi_a^\dagger[s] \,, & \l'=(s,+\hat e_a), \; s \in \Lambda;
	\\[2pt]
	\phi_a[s] \,, & \l'=(s,-\hat e_a), \; s \in \Lambda.
	\end {array}
    \right.
\label {eq:v3}
\end {equation}
``Gauge-scalar'' plaquettes (those containing both scalar and gauge links)
now reproduce the hopping terms in the scalar action (\ref {eq:Sscalar}).

By combining the above extensions of the underlying lattice,
one may, of course, consider a theory containing both fermions and scalars.
The net result is that
all bosonic single-trace gauge invariant observables
in the original lattice gauge theory with
adjoint representation scalars and/or fermions may be represented by
Wilson loops defined on a minimally extended lattice $\bar\Lambda$.

\subsection {Loop equations}
\label{sec:LE-parent}

Extending the previous derivation of loop equations to
lattice gauge theories containing adjoint representation matter fields
is straightforward.
We will use the following shorthand
for derivatives with respect to matter fields,
\begin {eqnarray}
    \delta^A_{s,a} &\equiv&
    (t^A)_{ij}\> \frac \delta{\delta \phi_a[s]_{ij}} \,,
\qquad
    \bar\delta^A_{s,a} \equiv
    (t^A)_{ij}\> \frac \delta{\delta \phi_a^\dagger[s]_{ij}} \,,
\label {eq:scalar delta}
\\[3pt]
    \delta^{A,b}_s &\equiv&
    (t^A)_{ij}\> \frac \delta{\delta \psi_b[s]_{ij}} \,,
\qquad
    \bar\delta^{A,b}_s \equiv
    (t^A)_{ij}\> \frac \delta{\delta \bar\psi_b[s]_{ij}} \,,
\label {eq:fermion delta}
\end {eqnarray}
so that
$
    \delta^A_{s,a} \, \phi_{a'}[s']
    = \delta_{ss'} \, \delta_{aa'} \> t^A
$,
{\em etc}.
Because our matter fields are not unitary,
these variations effectively replace a matter field by a 
Lie algebra generator, rather than (left) multiplying by a generator
[{\em c.f.} Eq.~(\ref {eq:delta})].
It remains true that the integral with the decoupled measure $d\mu_0$
of the variation of anything vanishes.
Defining
\begin {equation}
    \Delta \equiv
	{\textstyle \frac 1\Nc} \, e^{-S} \, 
	\Bigl\{
	{\sum_{\l \in\Lambda}}' \> \delta^A_\l \, e^S \, \delta^A_\l
	-
	\sum_{s\in\Lambda} \left[
	    \bar\delta^A_{s,a} \, e^S \, \delta^A_{s,a}
	    +
	    \delta^A_{s,a} \, e^S \, \bar\delta^A_{s,a}
	    +
	    \bar\delta^{A,b}_s \, e^S \, \delta^{A,b}_s
	    -
	    \delta^{A,b}_s \, e^S \, \bar\delta^{A,b}_s
	\right]
	\Bigr\} ,
\label {eq:Delta def}
\end {equation}
a natural generalization of the previous loop equation is simply
\begin {equation}
    0 = \left\langle \Delta \, {\cal O} \right\rangle 
\end {equation}
for any observable $\cal O$.
The motivation for the choice of signs in the definition (\ref {eq:Delta def})
is discussed below, after Eq.~(\ref{eq:loop eq 4}).

As just discussed,
single trace bosonic observables ({\em i.e.}, decorated Wilson loops)
may be associated with closed loops in the extended lattice $\bar\Lambda$.
If the observable $\cal O$ under consideration is
$
    W[C] \equiv \frac 1\Nc \, \tr \> v[C]
$
for some loop $C \in \bar\Lambda$,
then $\Delta \cal O$ will be a sum of
(a) terms proportional to $W[C]$ itself,
(b) decorated loops $W[C']$ where $C'$ is a deformation
of the loop $C$ produced by inserting a plaquette
(in the extended lattice $\bar\Lambda$), and
(c) products of loops $W[C']\,W[C'']$ (on the extended lattice)
produced by reconnecting $W[C]$ at self-intersections.
We discuss each of these three pieces in turn.

Terms proportional to $W[C]$ are generated in $\Delta W[C]$
when both link variations $\delta^A_\l$ act on the same link matrix in $W[C]$
(as seen earlier).
But such terms will now also be generated when a scalar variation
$\bar\delta^A_{s,a}$
(or $\delta^A_{s,a}$)
acts on the local potential part of the
scalar action.
This will bring down a factor of $\phi_a[s]$ (or $\phi_a^\dagger[s]$)
which will replace
an identical scalar field insertion in $W[C]$ which is removed by
the scalar variation $\delta^A_{s,a}$
(or $\bar\delta^A_{s,a}$)
acting directly on $W[C]$.
Similarly, the fermion variations $\delta^{A,b}_s$
(or $\bar\delta^{A,b}_s$),
when acting
on the mass term of the fermion action, will bring down a factor
of $\bar\psi_b[s]$
(or $\psi_b[s]$)
which will replace an identical insertion in $W[C]$
removed by the second fermion variation $\bar\delta^{A,b}_s$
(or $\delta^{A,b}_s$)
acting directly on $W[C]$.
The net result, after using large $\Nc$ factorization, is
\begin {eqnarray}
    \left\langle \Delta \, W[C] \right\rangle
    &=&
    \half \left(
	n_\l + V'[\left\langle\chi\right\rangle] \> n_{\rm s} +
	m \> n_{\rm f}
    \right) 
    \left\langle W[C] \right\rangle
\nonumber\\ && {}
    + \left\langle\Delta \, W[C] \right\rangle_{\rm deformation}
    + \left\langle\Delta \, W[C] \right\rangle_{\rm self-intersection}
    + O(1/\Nc^2)
    \,,
\label {eq:Delta Wa}
\end {eqnarray}
where $n_\l$ is the number of (ordinary) links contained in the loop $C$,
$n_{\rm s}$ is the number of scalar insertions
(both $\phi$'s and $\phi^\dagger$'s),
$n_{\rm f}$ is the number of fermion insertions
(both $\psi$'s and $\bar\psi$'s),
and $\chi \equiv \frac 1\Nc \sum_a \tr(\phi_a^\dagger[s] \, \phi_a[s])$.%
\footnote
    {
    The result (\ref {eq:Delta Wa}) assumes that the lattice is
    translationally invariant, so that $\langle \chi[s] \rangle$
    is independent of the site $s$.
    For such theories, note that the only
    dependence on the scalar potential $V[\chi]$ is via the single
    number $V'[\langle\chi\rangle]$;
    this is completely analogous to the large $\Nc$ universality of
    mixed adjoint-fundamental pure gauge actions
    \cite{Makeenko-Polikarpov,Samuel}.
    More generally, if the theory is not translationally invariant
    then $V'[\langle \chi \rangle] \, n_{\rm s}$
    should be replaced by $V'[\langle \chi[s] \rangle]$ summed over
    those sites at which scalar insertions appear in $W[C]$.
    }

As in the pure gauge theory,
deformations of the loop $C$ are produced by terms where one link variation
acts on the pure gauge action and the other link variation acts on $W[C]$.
In addition, there are now terms where one link variation acts
on the hopping terms of the matter field action
(and the other link variation acts on $W[C]$).
Such terms have the effect of inserting two matter fields
on either end of a link traversed by $C$.
Finally, there are terms where either a scalar or fermion variation acts on the
hopping terms of the matter field action
(and a scalar or fermion variation acts on $W[C]$).
These terms have the effect of moving a matter field insertion in $W[C]$
from its original site to some neighboring site.
All these terms may be regarded as deformations of the initial loop $C$
on the extended lattice $\bar\Lambda$ in which
a plaquette is inserted into the loop $C$.%
\footnote
    {
    For deformations arising from derivatives of matter field links,
    it would be more accurate to say that some link $\l$ is replaced
    by a ``staple'', {\em i.e.}, the three sides of a plaquette
    other than $\l$.
    We will continue to refer to all such deformations
    as plaquette insertions --- but the distinction is reflected
    in the presence of the $(\l \bar\l)^{-1}$ factor in Eq.~(\ref {eq:Delta W-deform}).
    }
The result may be expressed as
\begin {eqnarray}
    \left\langle\Delta \, W[C] \right\rangle_{\rm deformation}
    &=&
    -\sum_{\l \subset C} \sum_{p|\l \subset\partial p}
    \coeff 14 \, \betatilde_{\l,p}
    \left\{
    \left\langle W[(\overline {\partial p}) (\l \bar \l)^{-1} C]
    \right\rangle
    +
    s_{\l,p} \,
    \left\langle W[(\partial p) C]\right\rangle
    \right\} \!.
\label {eq:Delta W-deform}
\end {eqnarray}
The sum over plaquettes runs over all plaquettes in the extended lattice
$\bar\Lambda$ whose boundary includes the link $\l$.
Both $C$ and $\partial p$ are to be regarded
as starting with link $\l$ (oriented however it appears in $C$),
so that the concatenation of $\partial p$ with $C$ makes sense.
The factor of $(\l\bar\l)^{-1}$ in the first deformed loop
should be regarded as canceling the link $\l$ which begins $C$
and the link $\bar\l$ which ends $\overline {\partial p}$.
If directions of links are classified as `gauge'
({\em i.e.}, lying in the original lattice $\Lambda$),
`scalar', or `fermion',
then the plaquette weight $\betatilde_{\l,p}$ appearing in the result
(\ref {eq:Delta W-deform}) is
\begin {equation}
    \betatilde_{\l,p} \equiv
    \left\{
	\begin {array}{cl}
	    \beta_p/\Nc	\,, & \hbox {if $p$ is a `gauge-gauge' plaquette};\\
	    \eta\, \kappa /i \,,
			    & \hbox {if $p$ is a `gauge-fermion' plaquette};\\
	    \kappa	\,, & \hbox {if $p$ is a `gauge-scalar' plaquette};\\
	    0		\,, & \hbox {otherwise.}
	\end {array}
    \right.
\label {eq:beta tilde}
\end {equation}
For gauge-fermion plaquettes, the factor $\eta$
appearing in Eq.~(\ref {eq:beta tilde}) is the fermion flavor connection
$\eta[\l]$ if the link $\l$ being varied is a gauge link,
but if $\l$ is a fermion link, then $\eta = \eta[\l']$
where $\l'$ is the gauge link which precedes the forward directed
fermion link in $\partial p$
({\em i.e.}, the gauge link which runs from the $\psi$ to the $\bar\psi$).
The second term in the result (\ref {eq:Delta W-deform})
is present only for gauge links
with unitary connections
(for which a variation of $u[\l]$ also varies $u[\bar\l]$),
and not for matter field links with non-unitary connections.
For gauge links, the coefficient $s_{\l,p}$ is just a $\pm$ sign,
\begin {equation}
    s_{\l,p}
    \equiv
    \left\{
	\begin {array}{rl}
	    +1 ,\, & \hbox{if $\l$ is a gauge link and $p$ is a
	    		gauge-fermion plaquette};\\
	    -1 ,\, & \hbox{if $\l$ is a gauge link and $p$ is
	    		gauge-gauge or gauge-scalar};\\
	    0 ,\, & \hbox {if $\l$ is a scalar or fermion link}.
	\end {array}
    \right.
\label {eq:s_lp}
\end {equation}

Finally, the self-intersection terms involve a sum over all ways of
breaking the loop $C$ into two separate loops by reconnecting
each distinct pair of traversals of any multiply-traversed link.
But in the case of multiple traversals on matter field links,
only pairs of traversals in opposite directions contribute
(because $\psi$ and $\bar\psi$, or $\phi$ and $\phi^\dagger$, are distinct).
After using large $\Nc$ factorization,
the result is%
\footnote
    {
    Careful readers may note that
    double variations in $\Delta$, when acting on loops with fermion insertions,
    can generate terms which
    do not correspond to geometric self-intersections of the
    loop $C$ on the extended lattice $\bar\Lambda_{\rm f}$
    and which are not present in the result (\ref {eq:DelW-self}).
    This is a consequence of our having assigned each distinct integration
    variable to two links in $\bar\Lambda_{\rm f}$,
    as indicated in Eq.~(\ref {eq:v2}).
    However, the `missing' terms correspond to splitting the original
    bosonic observable into a product of two fermionic observables.
    The expectation value of these terms,
    after using large $\Nc$ factorization, will always vanish.
    }
\begin {equation}
    \left\langle\Delta \, W[C] \right\rangle_{\rm self-intersection}
    =
    \kern-5pt \sum_{\rm self-intersections} \kern-5pt
    I[\l] \,
    \left\langle W[C'] \right\rangle
    \left\langle W[C''] \right\rangle
    + O(1/\Nc^2) \,.
\label {eq:DelW-self}
\end {equation}
For parallel traversals of a gauge link $\l$,
$C = C'C''$ with loops
$C$, $C'$ and $C''$ all regarded as starting with link $\l$.
For antiparallel traversals, the loop $C$ is to be regarded as
$C = \l C' \bar\l C''$, with $\l$ positively oriented.
The self-intersection coefficient $I[\l]$ is
\begin {eqnarray}
    I[\l]
    &\equiv&
    \left\{
	\begin {array}{rl}
	+1,\, & \hbox {parallel traversals of a gauge link $\l$;}\\
	 0,\, & \hbox {parallel traversals of a scalar or fermion link $\l$;}\\
	-1,\, & \hbox {antiparallel traversals of link $\l$ (of any type);}\\
	\end {array}
    \right.
\nonumber\\[3pt] &\times&
    \left\{
	\begin {array}{rl}
	+1,\, & \hbox {if link $\l$ starts at a site in $\Lambda$;}\\
	-1,\, & \hbox {if link $\l$ starts at a site in $\widetilde\Lambda$.}
	\end {array}
    \right.
\label {eq:ICCC}
\end {eqnarray}

Combining these pieces yields a loop equation for expectation values
of single trace observables
on the extended lattice $\bar\Lambda$ which closely
resembles the result for a pure gauge theory,
\begin {eqnarray}
    \half \left(
	n_\l + V'[\left\langle\chi\right\rangle] \> n_{\rm s} +
	m \> n_{\rm f}
    \right) 
    \left\langle W[C] \right\rangle
    &=&
    \sum_{\l \subset C} \sum_{p|\l \subset\partial p}
    \coeff 14 \, \betatilde_{\l,p}
    \Bigl[
	\left\langle W[(\overline {\partial p}) (\l \bar \l)^{-1} C]
	\right\rangle
	+
	s_{\l,p} \,
	\left\langle W[(\partial p) C]\right\rangle
    \Bigr]
\nonumber\\ &-&
    \kern-5pt \sum_{\rm self-intersections} \kern-5pt
    I[\l] \,
    \left\langle W[C'] \right\rangle
    \left\langle W[C''] \right\rangle
    + O(1/\Nc^2) \,.
\label {eq:loop eq 4}
\end {eqnarray}

We will assume that $V'[\chi]$ is positive for non-negative arguments.%
\footnote
    {
    The hopping term of the scalar action (\ref {eq:Sscalar})
    differs from a lattice Laplacian by a local term proportional
    to $\phi[s]^\dagger\phi[s]$ --- this term has effectively been included in
    our scalar potential.
    Consequently, requiring positivity of $V'[\chi]$,
    even at $\chi = 0$, does not preclude the theory from being
    in a Higgs phase.
    }
We will also assume that the fermion mass $m$ is positive.
As long as all fermion species have a common mass, this is only
a matter of convention.
Consequently, the coefficient of $\langle W[C] \rangle$ on the left side
of the loop equation (\ref {eq:loop eq 4}) is strictly positive.
(The signs in the definition (\ref {eq:Delta def}) of the operator $\Delta$
were chosen so that this would be true.)
This means that one may iterate these loop equations
to generate the strong coupling expansion (in the large $\Nc$ limit)
for expectation values,%
\footnote
    {
    Starting with vanishing expectation values for all observables
    except
    $\frac 1\Nc \tr \, \bm 1$,
    $\frac 1\Nc \tr \> \bar\psi \psi$, and
    $\frac 1\Nc \tr \> \phi^\dagger \phi$.
    The appropriate initial values for $\langle \bar\psi\psi \rangle$
    and $\langle \phi^\dagger \phi \rangle$
    follow from the loop equation (\ref {eq:loop eq 4}) with
    all $\betatilde_{\l,p}$ set to zero.
    Specifically,
    $
	\langle \frac 1\Nc \tr \> \bar\psi_b \psi_{b'} \rangle
	=
	\delta_{bb'} \, m^{-1} + O(\betatilde_{\l,p})
    $
    and
    $
	\langle \frac 1\Nc \tr \> \phi_a^\dagger \phi_{a'} \rangle
	=
	\delta_{aa'} \, \langle \chi \rangle_0 / \Ns
	+ O(\betatilde_{\l,p})
    $,
    with $\langle \chi \rangle_0$ the (positive) root of
    $
	V'[\langle \chi \rangle] \, \langle \chi \rangle
	=
	1
    $.
    }
just like the pure gauge theory case.
And, once again, this expansion is guaranteed to be convergent
for sufficiently small values of $\betatilde_{\l,p}$.
In light of (\ref {eq:beta tilde}), this means both small hopping parameter
$\kappa$ (equivalent to large scalar or fermion mass)
as well as small $\beta_p/\Nc$ (or large 't Hooft coupling).
So, at least in the strong coupling/large mass phase of the theory,
the loop equations (\ref {eq:loop eq 4}) completely determine
the leading large $\Nc$ expectation values of single trace observables.

\subsection {Multi-loop connected correlators}
\label{sec:multi-loop-parent}

Applying exactly the same approach,
the equations for two-loop correlators 
can be obtained from
\begin{equation}
   \Big\langle \Delta \, \Big(\!\!
     \left(W[C_1]-\langle W[C_1]\rangle\right) 
     \left(W[C_2]-\langle W[C_2]\rangle\right) 
   \!\!\Big) \Big\rangle
   = 0 \,.
\label {eq:2-loop LE}
\end{equation}
This leads to the following equation for the
connected two-loop correlators:
\begin {eqnarray}
    && \half (
	n_\l + V'[\left\langle\chi\right\rangle] \> n_{\rm s} +
	m \> n_{\rm f}
    )\, 
    \dlangle W[C_1] W[C_2] \drangle
\nonumber\\[3pt]
    && \quad {}=
    \Biggl[\,
    \sum_{\l \subset C_1} \sum_{p|\l \subset\partial p} \!
    \coeff 14 \, \betatilde_{\l,p}
    \Bigl[
	\dlangle
	    W[(\overline {\partial p}) (\l \bar \l)^{-1} C_1] \, W[C_2]
	\drangle
	+
	s_{\l,p} \, \dlangle W[(\partial p) C_1] \, W[C_2] \drangle
    \Bigr]
    + (C_1 {\leftrightarrow} C_2)
    \Biggr]\!
\nonumber\\[2pt] && \quad {}
    -
    \Biggl[
    \sum_{{\scriptstyle {\rm self-intersections}
	\atop \vphantom{\tilde C}\scriptstyle (C_1)}}
    \kern-10pt
    I[\l] \,
    \Bigl[
	\dlangle W[C_1'] W[C_2]\drangle
	\left\langle W[C_1''] \right\rangle
	+
	\dlangle W[C_1'] W[C_2]\drangle
	\left\langle W[C_1''] \right\rangle
    \Bigr]
    + (C_1 {\leftrightarrow} C_2)
    \Biggr]
\nonumber \\[2pt] && \quad {}
    -
    \sum_{{\scriptstyle {\rm parallel\ gauge}
	\atop \scriptstyle {\rm mutual\ intersections}}
	\atop \scriptstyle \vphantom{\tilde C}(C_1,C_2)}
    \kern -10pt
	J[\l] \;
	\langle W[C_1 C_2] \rangle
    \; -
    \kern -10pt
    \sum_{{\scriptstyle {\rm anti-parallel}
	\atop \scriptstyle {\rm mutual\ intersections}}
	\atop \scriptstyle \vphantom{\tilde C}(C_1,C_2)}
    \kern -10pt
	K[\l] \;
	\langle W[C_1 (\l\bar\l)^{-1} C_2] \rangle 
\nonumber \\[2pt] && \quad {}
    +
    \sum_{{\scriptstyle {\rm parallel\ gauge}
	\atop \scriptstyle {\rm mutual\ intersections}}
	\atop \scriptstyle (\widetilde C_1,C_2)}
    \kern -10pt
	J[\l] \;
	\langle W[\widetilde C_1 C_2] \rangle
    \; +
    \kern -10pt
    \sum_{{\scriptstyle {\rm anti-parallel}
	\atop \scriptstyle {\rm mutual\ intersections}}
	\atop \scriptstyle (\widetilde C_1,C_2)}
    \kern -10pt
	K[\l] \;
	\langle W[\widetilde C_1 (\l\bar\l)^{-1} C_2] \rangle 
\nonumber \\[5pt] && \quad {}
    +
       O(1/\Nc^2)
    \,.
\label {eq:2-loop eq 2}
\end {eqnarray}
The numbers $n_\l$, $n_{\rm s}$, and $n_{\rm f}$ now denote the total
numbers of links, scalar insertions, and fermion insertions (respectively)
contained in both loops $C_1$ and $C_2$.
For self-intersection terms involving loop $C_1$,
if there is a parallel traversal
of some link $\l$ then
$C_1 = C_1'C_1''$ with loops $C_1$, $C_1'$ and $C_1''$ all regarded
as starting with $\l$, while for self-intersections with
antiparallel traversals, $C_1 = \l C_1' \bar\l C_1''$.
(And likewise for self-intersections involving $C_2$.)

As noted in footnote \ref{fn:Z2mirror}, the $Z_2$ image
$\widetilde C$ of any loop $C$ represents the same observable,
up to a minus sign, as does the loop $C$.
In order to describe all joinings of the two loops
which appear in the loop equation (\ref {eq:2-loop LE})
as geometric intersections, one must consider mutual intersections
between the two given loops in the extended lattice $\bar\Lambda$,
as well as mutual
intersections when either one of the loops is replaced by
its $Z_2$ image ({\em i.e.}, rigidly translated in a fermionic direction).
This is why two sets of mutual intersection terms appear
in the result (\ref {eq:2-loop eq 2}).
The `parallel gauge mutual intersection' sums run over mutual
intersections in which both loops traverse a gauge link
(not a matter field link) in the same direction;
both loops are to be regarded as starting with the intersection link $\l$.%
\footnote
    {
    If loops $C_1$ or $C_2$ multiply traverse a mutual intersection link $\l$,
    then each possible pairing of a traversal of $\l$ in $C_1$ with
    a traversal of $\l$ in $C_2$ generates a separate term in the
    parallel mutual intersection sum.
    Likewise, each possible pairing of a traversal of some link
    $\bar\l$ in $C_1$ with a traversal of $\l$ in $C_2$ generates
    a separate term in the antiparallel mutual intersection sum.
    }
The `anti-parallel mutual intersection' sums run over
mutual intersections in which loop $C_2$ traverses some link $\l$
while loop $C_1$ (or $\widetilde C_1$) traverses $\bar\l$;
$C_2$ is to be regarded as starting with $\l$
and $C_1$ (or $\widetilde C_1$) as ending with $\bar\l$.
The coefficient $J[\l]$ for parallel traversals of a gauge link $\l$ is
\begin {eqnarray}
    J[\l]
    = 
    \cases
	{
       +1, \, & \hbox{if $\l \in \Lambda$;}\cr
       -1, \, & \hbox{if $\l \in \widetilde\Lambda$,}
	}
\label {eq:JCC}
\end {eqnarray}
while the coefficient $K[\l]$ for antiparallel traversals of
a link $\l$ (of any type) is
\begin {eqnarray}
    K[\l]
    &=& 
    \cases{
    -1, \, & \hbox{if $\l$ is a gauge or scalar link;}\cr
    -1, \, & \hbox{if $\l$ is a forward-directed fermion ($\bar\psi$);}\cr
    +1, \, & \hbox{if $\l$ is a backward-directed fermion ($\psi$);}
    }
\nonumber\\[3pt] &\times&
    \left\{
	\begin {array}{rl}
	+1,\, & \hbox {if link $\l$ starts at a site in $\Lambda$;}\\
	-1,\, & \hbox {if link $\l$ starts at a site in $\widetilde\Lambda$.}
	\end {array}
    \right.
\label {eq:KCC}
\end {eqnarray}

The form of the result (\ref {eq:2-loop eq 2}) for connected correlators of
decorated loops is completely analogous to the result
(\ref {eq:2-loop eq}) for two-loop correlators in pure gauge theory;
the only differences are various fermionic minus signs and the absence of
deformation and intersection terms associated with parallel traversals
of matter field links.
In the same manner discussed previously, these connected correlator
loop equations may be
solved iteratively (starting with all two-loop correlators equal to zero)
to generate a strong coupling/large mass expansion with a non-zero
radius of convergence.
Consequently, these equations completely determine the
leading large $\Nc$ limit of two-loop connected correlators
(and hence the spectrum of particle masses%
\footnote
    {
    Although our geometric encoding of observables is, at the moment,
    restricted to bosonic observables,
    the extension to fermionic observables,
    discussed in Appendix~\ref{sec:Fermionic observables}, is straightforward.
    The above assertion (that loop equations for connected correlators
    determine the particle spectrum) is valid for fermionic as well
    as bosonic channels.
    }%
),
at least in the strong coupling/large mass phase of the theory.

\section{Orbifold theories}
\label{sec:Orbifolds}

\subsection {Orbifold projection}

Start with a $U(\Nc)$ gauge theory of the form discussed in the last section
[{\em c.f.} Eqs.~(\ref {eq:Smatter})--(\ref {eq:Sfermion})],
with $\Ns$ scalars and $\Nf$ fermions.
This will be referred to as the ``parent'' theory.
The global symmetry group of this theory
is $G = U(\Nc) \times U(\Ns) \times U(\Nf)$,
where the $U(\Nc)$ factor represents space-independent gauge transformations.
To make an orbifold projection, one chooses a subgroup $H$ of
this global symmetry group
and constructs a ``daughter'' theory by simply eliminating all degrees
of freedom in the parent theory which are not invariant under the chosen
subgroup $H$.
(A similar explanation of how to construct daughter theories
can be found in Refs.~\cite{Mithat1,Mithat2}.)

We will only consider projections based on Abelian subgroups,
and will specifically focus on cases where
\begin {equation}
    \Nc = k^d N \,,
\end {equation}
for some positive integers $k$ and $d$,
and where $H$ is a $(Z_k)^d$ subgroup of $G$
chosen so that the subgroup of the $U(k^d N)$ parent gauge group
which commutes with $H$ is $[U(N)]^{k^d}$.
This will be the gauge group of the daughter theory.
To specify the desired $(Z_k)^d$ subgroup of $G$,
it is sufficient to define the subgroup's $d$ independent generators ---
call them $\eta_\alpha$, $\alpha = 1, \cdots, d$.
Each generator will be the product of some gauge transformation
$\gamma_\alpha \in U(\Nc)$ times some non-gauge symmetry transformation
$h_\alpha \in U(\Ns) \times U(\Nf)$,
\begin {equation}
    \eta_\alpha = \gamma_\alpha \times h_\alpha \,.
\end {equation}
The gauge transformations $\{ \gamma_\alpha \}$,
regarded as $k^d N \times k^d N$ matrices,
generate a representation of $(Z_k)^d$ and may be chosen to be
\begin{equation}
    \gamma_\alpha
    =
    \underbrace{1_k \times\ldots}_{\alpha-1}{}\times \Omega \times
    \underbrace{1_k\times\ldots}_{d-\alpha}{} \times 1_N \,,
\label {eq:gamma def}
\end{equation}
where $1_N$ and $1_k$ are $N\times N$ and $k\times k$ unit matrices,
respectively, and
\begin {equation}
    \Omega \equiv {\rm diag}(\omega^0,\omega^1,\ldots,\omega^{k-1})
\end {equation}
with $\omega \equiv e^{2\pi i/k}$.
The factors $\{ h_\alpha \}$ must be elements of a
$U(1)^{\Ns+\Nf}$ maximal Abelian subgroup
of the non-gauge $U(\Ns) \times U(\Nf)$ symmetry group,
and each must be a $k$'th root of unity.
Hence one may write
\begin {equation}
    h_\alpha = e^{2\pi i \, r_\alpha / k} \,,
\end {equation}
where each $r_\alpha$ is a charge operator which assigns integer values
to matter fields in the theory (and zero to all gauge links).
(Different charge assignments will lead to differing daughter theories.)

If $\Phi$ denotes any variable (matter field or link variable)
in the parent theory,
all of which transform under the adjoint representation of the gauge group
and hence may be regarded as a $k^d N \times k^d N$ matrices,
then the action of the generator $\eta_\alpha$ on $\Phi$ is to transform
\begin {equation}
    \Phi \mapsto
    e^{2\pi i r_\alpha(\Phi) / k} \,
    \gamma_\alpha \, \Phi \, \gamma_\alpha^{-1} \,,
\end {equation}
where $r_\alpha(\Phi)$ is the value that the charge $r_\alpha$ assigns to the
variable $\Phi$.
Consequently, the net effect of the orbifold projection is the imposition
of the constraints
\begin {equation}
    \Phi = e^{2\pi i r_\alpha(\Phi)/k} \,
    \gamma_\alpha \, \Phi \, \gamma_\alpha^{-1} \,,\quad
    \alpha=1,\ldots,d \,,
\label {eq:constraint}
\end {equation}
on each adjoint representation variable $\Phi$.

At this point, it is useful to introduce the terminology of ``theory space''
\footnote{
	The term ``theory space'' was introduced in \cite{ACG}.
	Other often used names are ``quiver diagrams'' and ``moose diagrams''.}
which provides the natural ``habitat'' for discussing
the field content of the daughter theory.
Theory space is a graph, denoted $T$, consisting of points and
(directed) bonds.%
\footnote
    {
    We are avoiding use of the words ``sites'' and ``links''
    to describe the theory space graph,
    to prevent confusion with the previous use of sites and links
    in reference to the spacetime lattice.
    }
Each point denotes a $U(N)$ factor of the daughter theory gauge group.
Each bond represents a matter field transforming
under the fundamental representation of the gauge group factor at the
originating end of the bond, and under the anti-fundamental
representation of the gauge group factor at the final end of the bond
(and transforming as a singlet under all other gauge group factors);
these are termed `bifundamentals'.
In our chosen case of a $(Z_k)^d$ orbifold,
we have a theory space with $k^d$ points
which may be regarded as forming a regular, periodic
lattice discretization of a $d$-dimensional torus.
Theory space points [or associated $U(N)$ factors of the daughter gauge group]
may be labeled by a $d$-dimensional vector $\j$ whose components
are integers running from 0 to $k{-}1$ (modulo $k$).

Let $\r = \{ r_\alpha(\Phi) \}$ denote the vector of charge assignments
for a particular field $\Phi$.
The link variables $u[\l]$ must all have vanishing charge vectors,
since they do not transform under the non-gauge symmetries $h_\alpha$.
Consequently, for link variables, the orbifold projection constraints
(\ref {eq:constraint})
imply that each $k^d N \times k^d N$ unitary link
matrix must be block-diagonal with $k^d$ independent blocks,
each of which is an $N\times N$ unitary matrix.
Each block is the gauge connection, in the daughter theory,
for one of the $U(N)$ factors of the $U(N)^{k^d}$ daughter gauge group;
the individual blocks may be labeled as $u^\j[\l]$ for $\j \in T$.

Each parent matter field, after the orbifold projection
(\ref {eq:constraint}), generates $k^d$ bifundamental fields in the
daughter theory.
For a matter field with charge vector $\r$,
these bifundamental fields may be represented by
bonds in the theory space connecting each point $\j$ with point $\j+\r$.
More explicitly, the variables of the daughter theory are
the unitary link variables $u^{\j}[\l] \in U(N)_\j$
belonging to each of the $k^d$ gauge group factors,
together with $\Ns \, k^d$ complex scalar bifundamentals $\phi^\j_a[s]$
and $\Nf \, k^d$ pairs of Grassmann bifundamentals
$(\psi^\j_b[s],\bar \psi^\j_b[s])$
on each site of the (physical) lattice.
The gauge transformation properties of the matter variables may be summarized as
\begin{eqnarray}
	\phi^{\j}_{a}[s]
    &\quad:\quad&
	( \Yfund_{\j}, \overline{\Yfund}_{\j+ \r_{a} } ) \,,
\\
	{\phi}^{\j}_{a}[s]\rlap{$^\dagger$}
    &\quad:\quad&
	( \overline{\Yfund}_{\j}, \Yfund_{\j+ \r_{a}}) \,,
\\
	\psi^{\j}_{b}[s]
    &\quad:\quad&
	( \Yfund_{\j}, \overline{\Yfund}_{\j+ \r^{b} } ) \,,
\\
	\bar{\psi}^{\j}_{b}[s]
    &\quad:\quad&
	( \overline{\Yfund}_{\j}, \Yfund_{\j + \r^{b}}) \,,
\end{eqnarray}
where $\r_a$ ($a = 1,\cdots,\Ns$)
is the vector of charge assignments for the parent
scalar field $\phi_a[s]$, and $\r^b$ ($b = 1,\cdots,\Nf$)
is the corresponding charge vector for the parent fermion field $\psi_b[s]$;
note that these charge assignments are independent for each matter field.

\begin{FIGURE}[t]
    {
    \parbox[c]{\textwidth} 
	{
	\begin{center}
	\includegraphics[width=0.9\textwidth]{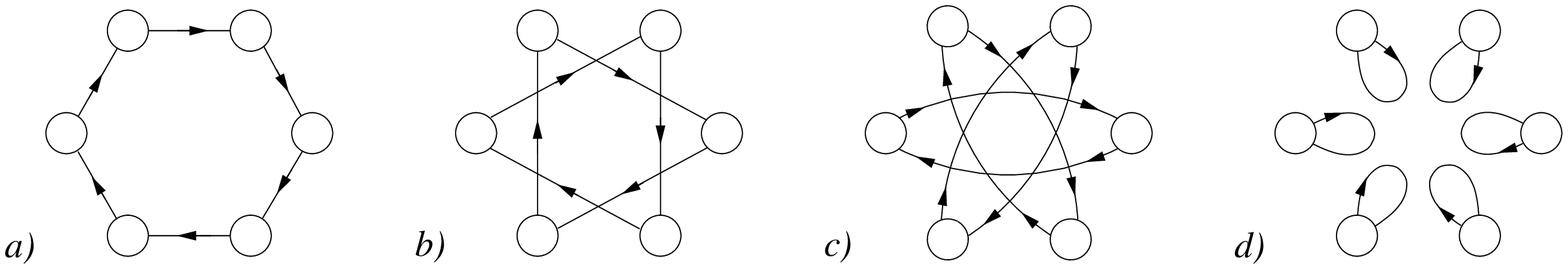}
	\caption
	    {
	    Theory space graphs obtained by applying a $Z_6$ orbifold
	    projection to a $U(6N)$ gauge theory containing a
	    single adjoint representation fermion.
	    The different graphs result from differing $r$ charge
	    assignments for the scalar field;
	    graphs $(a)$--$(d)$ correspond to $r = 1, 2, 3$ and 0,
	    respectively.
	    (Each bond represents a portion of the fermion field
	    $\psi[s]$ which survives the orbifold projection;
	    projections of $\bar\psi[s]$ correspond to reversing all arrows.)
	    }
	\label {fig:orbifolds1}
	\end{center}
	}
    }
\end{FIGURE}

Figure \ref {fig:orbifolds1} illustrates the resulting theory space
for $Z_6$ orbifold projections ({\em i.e.}, $d =1$ and $k=6$)
in a theory with one adjoint fermion
and differing $\r$ charge assignments.
These orbifold projections involve a single gauge transformation
$\gamma \in U(6N)$ [{\em c.f.} Eq.~(\ref {eq:gamma def})]
which has the form
\begin{equation}
    \gamma= \left(
	    \begin{array}{rrcr}
		    1_{N \times N}
		\\
		    & \omega \, 1_{N \times N} &
		\\
		    & & \ddots
		\\
		    & & & \omega^{5} \, 1_{N \times N}
	    \end{array} 
	    \right)
\end{equation}
where $\omega = e^{2 \pi i/6}$.
The link variables of the parent theory are $U(6 N)$ matrices.
Since link variables have vanishing $\r$ charge,
the effect of the projection (\ref {eq:constraint}) is to restrict
each link variable $u[\l]$ to be block-diagonal, with six blocks
each of which is an $N \times N$ unitary matrix;
these are precisely the daughter link variables $u^j[\l]$, $j = 1,\cdots,6$.

Since $d=1$, the charge vector assigned to the fermion field $\psi[s]$
is only a single integer $r$.
If the fermion is assigned vanishing charge, $r=0$,
then the orbifold projection restricts these variables to be
block diagonal, just like the link variables.
In this case, illustrated in Fig.~\ref {fig:orbifolds1}d,
the net effect is to reduce the $U(6N)$ parent theory to six
decoupled copies of a $U(N)$ gauge theory with one adjoint fermion
({\em i.e.}, the same theory as the parent except for the smaller
gauge group).%
\footnote
    {
    If one replaces the parent fermion by a scalar field
    then, in the daughter theory,
    the quartic self-interactions couple the six different scalars
    so that the daughter theory is no longer a product of six
    independent theories.
    }
If the charge $r$ assigned to the fermion is non-vanishing,
then the effect of the projection (\ref {eq:constraint}) is to restrict
these variables to a form in which each variable has six $N \times N$
non-zero blocks
that form a diagonal stripe displaced from the principle diagonal
by $r$ (mod 6) steps.%
\footnote
    {
    If the parent field is divided into 36 blocks (each $N \times N$),
    labeled $(j,j')$ with $j,j'=1,\cdots,6$,
    then the orbifold projection eliminates all blocks except those with
    $j'-j = r \bmod 6$.
    }
As Fig.~\ref {fig:orbifolds1} illustrates, if
$r=1 \bmod 6$, one obtains a daughter theory with bifundamental
fermions transforming under adjacent $U(N)$ gauge group factors.
There is a manifest $Z_6$ discrete symmetry which cyclically permutes
the six gauge group factors.
For $r=2 \bmod 6$, one obtains two decoupled copies of a
$U(N)^3$ gauge theory in which a trio of bifundamental fermions
connect the factors.
For $r=3 \bmod 6$, one has
three decoupled $U(N)^2$ gauge theories, each containing a pair
of bifundamental fermions.
The graphs for $r=4$ or 5 (mod 6) are the same as those for $r=2$
or 1, respectively, with the directions of arrows reversed;
{\em i.e.}, the daughter fermions are in conjugate representations.

\begin{FIGURE}[t]
    {
    \parbox[c]{\textwidth} 
	{
	\begin{center}
	\includegraphics[width=0.16\textwidth]{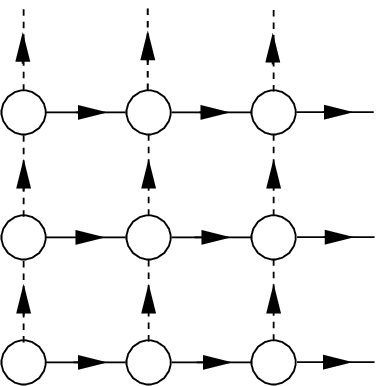}
	\caption
	    {
	    Theory space graph obtained from a $(Z_3)^2$ orbifold
	    projection on a $U(9N)$ gauge theory containing
	    one adjoint scalar and one adjoint fermion.
	    The ${\r}$ charge vector for the scalar is $(0,1)$
	    while that of the fermion is $(1,0)$.
	    The graph is periodic in both directions;
	    the dangling bonds at the top and right
	    edges should be understood as wrapping around and
	    connecting with the corresponding points along the
	    bottom and left edges, respectively.
	    Bonds drawn with solid lines represent bifundamental
	    fermions, while dashed bonds represent bifundamental scalars.
	    }
	\label {fig:orbifolds2}
	\end{center}
	}
    }
\end{FIGURE}

Figure \ref {fig:orbifolds2} illustrates the case of a $(Z_3)^2$
orbifold projection on a $U(9 N)$ gauge theory with one adjoint scalar
and one adjoint fermion.
We have chosen the  ${\r}$ charges to be  $(0,1)$ for the scalar,
and $(1,0)$ for the fermion.
All variables are now subjected to two constraints of the form
(\ref {eq:constraint}).
If each (adjoint representation) parent variable is divided into
a $9 \times 9$ array of blocks, each of which is $N \times N$,
then only 9 blocks from each variable will satisfy both constraints.
The daughter theory has a $U(N)^9$ gauge group,
9 bifundamental scalars, and 9 bifundamental fermions
transforming as indicated in the theory space graph
of Fig.~\ref {fig:orbifolds2}.
The graph should be regarded as periodic in both directions
so that it is invariant under discrete translations.
This reflects the fact that the daughter theory has a $(Z_3)^2$
discrete global symmetry which permutes the different gauge group factors.

Returning now to the discussion of our general class of $(Z_k)^d$ orbifolds,
we will define the daughter theory action $S^{(\d)}$ to be the result
of replacing every variable in the parent theory action
(\ref {eq:Smatter})--(\ref {eq:Sfermion})
by its orbifold projection,
and then rescaling the action by a factor of $N/\Nc = k^{-d}$.
Including this rescaling will be necessary to make the
daughter theory loop equations isomorphic to those of the parent theory.
The resulting action of the daughter theory is
\begin{equation}
    S^{(\d)}
    =
    S_{\rm gauge}^{(\d)} +
    S_{\rm scalar}^{(\d)} +
    S_{\rm fermion}^{(\d)} \,,
\label {eq:orbS}
\end{equation}
with
\begin {equation}
    S_{\rm gauge}^{(\d)}
    \equiv
    \sum_{{\j}\in T } \>
    {\sum_{p \in \Lambda}}' \> \beta_p^{(\d)} \>
    \Re\, \tr \, u^{\j} [\partial p] \,,
\label {eq:orbS-YM}
\end {equation}
and
\begin {equation}
    {\beta_p^{(\d)} \over N} \equiv {\beta_p \over \Nc} \,.
\label {eq:equal tHooft}
\label{eq:coupling rescaling}
\end {equation}
Note that this gauge action involves a sum over each point in the
theory space $T$ ({\em i.e.}, a sum over each $U(N)$ gauge group factor).
The condition (\ref {eq:equal tHooft}) is equivalent to the requirement
that the 't Hooft couplings ($g^2 N$) coincide in the parent and daughter
theories.
The scalar action in the daughter theory is
\begin {eqnarray}
    S_{\rm scalar}^{(\d)}
    &=&
    N \Biggl\{ \>
    \sum_{\l=\langle ss' \rangle \in\Lambda} \>
    \sum_{a=1}^{\Ns} \>
    \sum_{{\j} \in T} \>
	\half \, \kappa \> \tr \!
	\left(
	    {\phi}_a^{\j}[s]^\dagger u^{\j}[\l] \,
	    \phi_a^{\j}[s'] \, u^{{\j+ \r_{a}}}[\bar \l]
	\right)
\nonumber\\ && \qquad {}
    -
    \sum_{s\in\Lambda} \>
	\Nc \, V\!\Bigl[
	\sum_{a=1}^{\Ns} \>
	\sum_{\j\in T}
	\smash{\tr \over \Nc} \!\!
	\left({\phi}_a^{\j}[s]^\dagger \phi_a^{\j}[s]\right)
	\Bigr]
    \Biggr\}
    \,.
\label {eq:orbSscalar}
\end {eqnarray}
This generalization of the parent scalar action
(\ref {eq:Sscalar}) describes a set of $k^d$ scalars which
are in either adjoint or bifundamental representations,
depending on whether their $\r$ charges vanish or are non-zero.
Finally, the fermion action in the daughter theory is
\begin {equation}
    S_{\rm fermion}^{(\d)}
    =
    N
    \sum_{b=1}^{\Nf}
    \sum_{{\j} \in T}
    \Biggl\{ \>
    \sum_{\l=\langle ss' \rangle \in\Lambda} \kern-5pt
    \coeff 1 {2i} \, \kappa \>
	\tr \!
	\left(
	    \bar\psi^{\j}_b[s] \, \eta[\l] \, u^{\j}[\l] \, 
	    \psi^{\j}_b[s'] \, u^{\j + \r^{b}}[\bar \l]
	\right)
    -
    \sum_{s\in\Lambda} \, 
	m \>\tr \! \left(\bar\psi^{\j}_b[s] \, \psi^{\j}_b[s]\right)
    \Biggr\}
    \,.
\label {eq:orbSfermion}
\end {equation}

In addition to whatever discrete translation and rotational symmetries are
possessed by the Euclidean lattice $\Lambda$,
the daughter theory action (\ref {eq:orbS}),
and associated integration measure,
are invariant under independent $U(N)$ gauge transformations in each of the
$k^d$ gauge group factors.
The daughter theory is also invariant under a $(Z_k)^d$ global symmetry
which permutes the different gauge group factors
and fields of the daughter theory in the manner dictated by the
discrete translation symmetry of the periodic theory space graph $T$.
Finally, the daughter theory is invariant under whatever subgroup
of the global $U(\Ns) \times U(\Nf)$ flavor rotation group commutes
with the non-gauge transformations $\{ h_\alpha \}$ used to define
the orbifold projection.%
\footnote
    {
    Retaining the full $U(\Ns)$ symmetry in the daughter
    theory requires that all scalars have the same $\r$ charge.
    Retaining $U(\Nf)$ symmetry likewise requires that all fermions
    have a common $\r$ charge.
    If distinct $\r$ charges are assigned to different flavors
    of fermions or scalars, then the global flavor symmetry of the
    daughter theory will be a smaller subgroup of $U(\Ns) \times U(\Nf)$.
    }

As already illustrated by the daughter theory action (\ref {eq:orbS}),
the orbifold projection connecting parent and daughter theories
generates a natural mapping between observables of the 
parent theory and the subclass of observables of the daughter
theory which are invariant under the $(Z_k)^d$ global symmetry.
To see this explicitly, consider first an ordinary Wilson loop $W[C]$
under, for simplicity, a $Z_k$ orbifold projection.
If the link variables composing the Wilson loop in the parent theory
are replaced by their orbifold projections,
then the Wilson loop operator
becomes an average of loop operators (for the same contour)
in each of the $k$ different $U(N)$
gauge group factors of the daughter theory,
\begin {equation}
    W[C] \equiv \frac {1}{kN} \> \tr \> u[C]
    \longrightarrow
    \frac {1}{k} \sum_{j=1}^{k} \, \frac {1}{N} \, \tr \> u^{j}[C]
    \equiv W_\d[C] \,.
\label {eq:map1W[C]}
\end {equation}
The trace on the left side of the map involves $kN \times kN$ matrices,
while that on the right involves $N \times N$ matrices.
As indicated here, we will use $W_\d[C]$ to denote daughter theory
Wilson loops which are averaged over daughter theory gauge group factors
(or equivalently, over all points of the theory space $T$).

Now consider of a decorated Wilson loop
({\em i.e.}, one with matter field insertions)
such as, for example,
$
    W[\Gamma_1, \Gamma_2]_{b \bar b}
    \equiv
    \frac {1}{kN} \> \tr \>\psi_b[s_1] u[\Gamma_1] \bar\psi_b[s_2] u[\Gamma_2]
$.
If all variables are replaced by their orbifold projections, then
\begin {eqnarray}
    && W[\Gamma_1, \Gamma_2]_{b \bar b}
    \equiv
    \frac {1}{kN} \, 
    \tr \; \psi_{b}[s_1] \, u[\Gamma_1] \, \bar\psi_{b}[s_2] \, u[\Gamma_2]
\nonumber\\ && \qquad \downarrow
\label {eq:map2W[C]}
\\ 
    && W_\d[\Gamma_1,\Gamma_2]_{b\bar b}
    \equiv
    \frac {1}{k}
    \sum_{j=1}^{k} \, \frac {1}{N} \, \tr \;
    \psi_{b}^{j}[s_1] \, u^{{j + r}^{b}}[\Gamma_1] \,
    \bar\psi_{b}^{j}[s_2] \, u^{j}[\Gamma_2] \,,
\nonumber
\end {eqnarray}
where $r^b$ is the $\r$ charge assigned to fermion $\psi_b$.
If the charge $r^b$ is non-zero then the daughter fermion $\psi^j_b$
is in a bifundamental representation, which means it transforms
under different gauge group factors on the left and right.
This is why the appropriate gauge connections (emerging directly
from the orbifold projection) involve $u^j$ when acting on the left
of $\psi^j_b$, and $u^{j+r^b}$ when acting on the right.
The result is a gauge invariant operator in the daughter theory
(as it must be) which, once again, is averaged over all ``starting points''
in theory space and thus is invariant under the $Z_k$ global symmetry
of theory space.

More generally, if ${\cal O}$ is any operator of the
parent theory which is both gauge invariant and invariant under
the non-gauge symmetry transformations $\{ h_a \}$ used in defining the
orbifold projection,
then the projection will map this operator into an operator ${\cal O}_\d$
in the daughter theory which is both gauge invariant and invariant
under the global $(Z_k)^d$ translation symmetry of theory space.%
\footnote
    {
    If the operator $\cal O$ transforms non-trivially
    (and irreducibly) under the $\{ h_a \}$ non-gauge symmetries,
    then it maps to zero under the orbifold projection.
    A simple example is $\tr \, \phi^n$ in a $Z_k$ orbifold
    with a scalar field $\phi$.
    If $\phi$ has non-zero $\r$ charge, then after the orbifold projection
    $\phi^n$ will be block off-diagonal and $\tr \, \phi^n$ will vanish
    unless $n \, r$ is divisible by $k$.
    The condition that $\cal O$ be invariant under the
    $\{ h_a \}$ non-gauge symmetries amounts to the requirement
    that the $\r$-charges of all matter field insertions sum to zero.
    }
As a final example, consider a Wilson loop decorated by any number of
fermion (or antifermion) insertions in a multi-flavor theory.
Under a general $(Z_k)^d$ orbifold projection,
\begin {eqnarray}
    &&
	W[\Gamma_1, \Gamma_2, \cdots, \Gamma_K]_{b_1 \bar b_2\cdots b_K}
    = 
	\frac {1}{k^dN} \> \tr
	\left(
	    \psi_{b_1}[s_1] \, u[\Gamma_1] \,
	    \bar\psi_{b_2}[s_2] \, u[\Gamma_2] \cdots u[\Gamma_{K-1}] \,
	    \psi_{b_K}[s_K] \, u[\Gamma_K]
	    ^{\vphantom{\dagger}}
	\right)
\nonumber\\ && \qquad \downarrow \vphantom {\Big|}
\label {eq:map3W[C]}
\\ &&
	W_\d[\Gamma_1, \Gamma_2, \cdots, \Gamma_K]_{b_1 \bar b_2\cdots b_K}
	\equiv
	\frac{1}{k^d} \sum_{{\j \in T} }\frac {1}{N} \>
	\tr \left(
	    \psi_{b_1}^{\j}[s_1] \, u^{\j + \r^{b_1}}[\Gamma_1] \,
	    \bar\psi_{b_2}^{\j + \r^{b_1} }[s_2] \, 
	   u^{\j + \r^{b_1} - \r^{b_2}}[\Gamma_2]
	   \cdots {}
	   \right.
\nonumber\\ && \kern 3in
	   \left.
	   {} \cdots u^{\j - \r^{b_K} }[\Gamma_{K-1}] \,
	   \psi_{b_K}^{\j - \r^{b_K}}[s_K] \,
	   u^{\j}[\Gamma_K] ^{\vphantom{\dagger}}
	\right) ,
\nonumber
\end {eqnarray} 
provided the sum of $\r$ charges of all the fermion insertions vanish,
$\r^{b_1} - \r^{b_2} + \cdots + \r^{b_K} = 0$ (mod $k$).
(Otherwise, the operator maps to zero under the orbifold projection.)
Associating each variable with a point or bond in theory space,
as discussed earlier, this condition is the same as the requirement
that the path {\em in theory space} traversed by a single-trace
operator must be closed.
Note that given an arbitrary starting point $\j$ in theory space,
the transformation properties of each variable in the parent operator
uniquely determine the path in theory space associated with the
daughter operator.
The starting point $\j$ is averaged over all points in theory space,
thereby explicitly constructing a $(Z_k)^d$ invariant result.

Once again, instead of displaying explicitly the path segments
and insertions in decorated loops
[as in Eqs.~(\ref {eq:map2W[C]}) and (\ref {eq:map3W[C]})],
one may instead associate every such decorated loop with a closed contour $C$
in the extended lattice $\bar\Lambda$,%
\footnote
    {
    As stated earlier, we are assuming for the moment that all operators
    are bosonic.
    }
and write the operator mapping of arbitrary (bosonic) single-trace operators
in the trivial form
\begin {equation}
    W[C] \longrightarrow W_\d[C] \,, \qquad C \in \bar\Lambda \,.
\end {equation}
The essential point is that any closed path $C$ in the extended lattice
uniquely identifies both the associated operator $W[C]$ in the parent theory,
and the corresponding $(Z_k)^d$ invariant operator $W_\d[C]$
in the daughter theory.

\subsection{Loop equations in daughter orbifold theories}

The previous treatment of loop equations in theories with adjoint
matter fields may be generalized to daughter orbifold theories in a
straightforward fashion.
The operator $\Delta$ [{\em c.f.}, Eq.~(\ref {eq:Delta def})]
which generated our previous loop equations must
merely be redefined to include a sum over all points in theory space,
\begin {eqnarray}
    \Delta
    &\equiv&
    {\textstyle \frac 1N} \, e^{-S^{(\d)}} \, {\sum_{{\j} \in {T}}}
    \biggl\{	
	{\sum_{\l \in\Lambda}}' \>
	\delta^{A,{\j}}_\l \, e^{S^{(\d)}} \, \delta^{A,{\j}}_\l
\label {eq:OrbDelta def}
\\ && \qquad\qquad {}
	-
	\sum_{s\in\Lambda}
	\left[
	    \bar\delta^{A,{\j}}_{s,a} \, e^{S^{(\d)}} \, \delta^{A,{\j}}_{s,a}
	    +
	    \delta^{A,{\j}}_{s,a} \, e^{S^{(\d)}} \, \bar\delta^{A,{\j}}_{s,a}
	    +
	    \bar\delta^{A,\j,b}_{s} \, e^{S^{(\d)}} \, \delta^{A,{\j},b}_{s}
	    -
	    \delta^{A,{\j},b}_{s} \, e^{S^{(\d)}} \, \bar\delta^{A,{\j},b}_{s}
	\right]
    \biggr\} .
\nonumber
\end {eqnarray}
Here $\delta^{A,\j}_\l$ is the link variation previously defined in
Eq.~(\ref{eq:delta}), but now acting specifically on the link variable
$u^\j[\l]$.
Similarly, $\delta^{A,\j}_{s,a}$ is a scalar variation as
defined in Eq.~(\ref {eq:scalar delta}) but now acting on $\phi^\j_a[s]$,
and $\delta^{A,\j,b}_s$ is the fermion variation as
defined in Eq.~(\ref {eq:fermion delta}) but now acting on $\psi^\j_b[s]$,
{\em etc}.
The integral of any variation still vanishes, so the loop equation
for any observable $\cal O$ in the daughter theory may once again be written as
\begin {equation}
    0 = \left\langle \Delta \, {\cal O} \right\rangle  \,.
\end {equation}

For any closed contour $C$ in the extended lattice $\bar\Lambda$,
let $W[C]$ denote the associated single-trace decorated Wilson loop
in the parent theory, and $W_\d[C]$ the corresponding single-trace
$(Z_k)^d$ invariant decorated Wilson loop in the daughter theory.
Just as in the parent theory,
the daughter theory loop equation for $W_\d[C]$
will involve a sum of three types of terms:
terms proportional to $\langle W_\d[C] \rangle$,
terms involving single plaquette
deformations (in the extended lattice) of the contour $C$,
and self-intersection terms.

Terms proportional to $\langle W_\d[C] \rangle$ are generated when
both link variations
$\delta^{A,\j}_\l$ in the operator $\Delta$ act on the same link variable
$u^\j[\l]$ present in (some piece of) $W_\d[C]$.
Such terms are also generated when one matter field variation
acts on the local part of the matter field action
(the fermion mass term or the scalar potential term)
and the other variation acts on an insertion of the conjugate matter field
in $W_\d[C]$.
Due to the inclusion of a sum over all theory space points in the
definition (\ref {eq:OrbDelta def}), the resulting contribution from
each gauge link, scalar, or fermion insertion in $W_\d[C]$ is independent
of the theory space index on the variable.
The rescaling of the fermion mass and scalar potential terms in the
daughter theory action (relative to the parent action)
by a factor of $N/\Nc$  is exactly what is needed so that,
by construction, the resulting coefficient
of $\langle W_\d[C]\rangle$ is the same
as in the parent theory.
In other words,
\begin {eqnarray}
    \left\langle \Delta \, W_\d[C] \right\rangle
    &=&
    \half \left(
	n_\l + V'[\left\langle\chi\right\rangle] \> n_{\rm s} +
	m \, n_{\rm f}
    \right) 
    \left\langle W_\d[C] \right\rangle
\nonumber\\ && {}
    + \left\langle\Delta \, W_\d[C] \right\rangle_{\rm deformation}
    + \left\langle\Delta \, W_\d[C] \right\rangle_{\rm self-intersection}
    + O(1/N^2)
    \,,
\label {eq:orbDelta Wd}
\end {eqnarray}
where
$
    \chi \equiv
    \sum_{\j,a} {\tr\over\Nc}(\phi^\j_a[s]^\dagger \, \phi^\j_a[s])
$,
and $n_\l$ denotes the number of gauge links in the contour $C$,
$n_{\rm s}$ the total number of scalar insertions,
and $n_{\rm f}$ the total number of fermion insertions.
 
Deformations of the loop $C$ are produced whenever one link variation 
$\delta^{a,{\j}}_\l$ acts on the gauge action $S^{(\d)}_{\rm gauge}$
and the other variation
acts on a gauge link $u^\j[\l]$ present in $W_\d[C]$,
or when a matter field variation acts on the hopping terms in the
action and the conjugate variation acts on a matter field insertion
in $W_\d[C]$.
As described earlier, all of these terms may be regarded
as plaquette deformations in the extended lattice $\bar\Lambda$;
the fact that all variables now carry an additional theory space label
$\j$ makes no difference.
The result may be written in the form
\begin {equation}
    \left\langle\Delta \, W_\d[C] \right\rangle_{\rm deformation}
    =
    -\sum_{\l \subset C} \sum_{p|\l \subset\partial p}
    \coeff 14 \, \betatilde^{(\d)}_{\l,p}
    \left\{
	\left\langle W_\d[(\overline {\partial p}) (\l \bar \l)^{-1} C]
	\right\rangle
	+
	s_{\l,p} \,
	\left\langle W_\d[(\partial p) C]\right\rangle
    \right\} ,
\label {eq:orbDelta W-deform}
\end {equation}
where the coefficient $\betatilde_{\l,p}^{(\d)}$ equals $\beta_p^{(\d)}/N$
when $p$ is a `gauge-gauge' plaquette, and is otherwise the same
as $\betatilde_{\l,p}$ as defined in Eq.~(\ref {eq:beta tilde}).
[And $s_{\l,p}$ is the same coefficient defined previously in
Eq.~(\ref {eq:s_lp}).]
Hence, given the relation (\ref {eq:equal tHooft}) between parent and
daughter gauge couplings, this result
coincides precisely with the corresponding deformation term
(\ref {eq:Delta W-deform})
in the parent theory.

The final contributions to the loop equation for $W_\d[C]$ are
self-intersection terms produced by double variations in $\Delta$ 
acting on multiply traversed links.
In these terms, there {\em is} a potential difference
between parent and daughter theories.
When a decorated Wilson loop is represented as a closed contour
in the extended lattice,
self-intersection terms in the parent theory loop equation may be regarded
as geometric;
every pair of traversals of any given link
(in opposite directions for matter field links,
and either direction for gauge links)
generates a self-intersection contribution.
In the daughter theory, analogous self-intersection terms are only present
when the two traversals of the given link represent variables
with the same theory space label $\j$.
This follows directly from the structure of the
operator $\Delta$ (\ref {eq:OrbDelta def}):
gauge invariance dictates that both variations in each term
act on variables at the same place in theory space.

As a concrete example,
consider the parent theory observable
\begin {equation}
    {\cal O}
    \equiv
    {\tr\over \Nc} \left(
    \phi_a[s] \, u[C_1] \,
    \phi_a[s] \, u[C_2] \,
    \phi_a[s]^\dagger \, u[C_3] \,
    \phi_a[s]^\dagger \, u[C_4] \right) ,
\label {eq:O example}
\end {equation}
containing four scalar field insertions all at the same site $s$.
($C_1, \cdots, C_4$ are all closed loops
in the physical lattice $\Lambda$ which begin at site $s$.)
The corresponding daughter theory observable is
\begin {equation}
    {\cal O}_\d
    \equiv
    {1 \over k^d}
    \sum_{\j\in T} \>
    {\tr\over N}
    \left(
    \phi^\j_a[s] \, u^{\j+\r_a}[C_1] \,
    \phi^{\j+\r_a}_a[s] \, u^{\j+2\r_a}[C_2] \,
    \phi^{\j+\r_a}_a[s]^\dagger \, u^{\j+\r_a}[C_3] \,
    \phi^{\j}_a[s]^\dagger \, u^\j[C_4] \right) .
\end {equation}
In the loop equation for $\cal O$,
the self-intersection terms
(after using large $\Nc$ factorization)
generated by double variations of the scalar fields are
\begin {eqnarray}
    \langle {\cal O} \rangle_{\rm self-intersection}
    &=&
    - \Bigl\langle
	{\tr\over \Nc} \>
	u[C_1] \, \phi_a[s] \, u[C_2] \, \phi_a[s]^\dagger \, u[C_3]
    \Bigr\rangle
    \Bigl\langle
	{\tr\over \Nc} \>
	u[C_4] \Bigr\rangle
\nonumber\\ && {}
    - \Bigl\langle
	{\tr\over \Nc} \>
	u[C_1] \, \phi_a[s] \, u[C_2]
    \Bigr\rangle
    \Bigl\langle
	{\tr\over \Nc} \>
	u[C_3] \, \phi_a[s]^\dagger \, u[C_4]
    \Bigr\rangle
\nonumber\\ && {}
    - \Bigl\langle
	{\tr\over \Nc} \>
	u[C_2] \, \phi_a[s]^\dagger \, u[C_3]
    \Bigr\rangle
    \Bigl\langle
	{\tr\over \Nc} \>
	u[C_4] \, \phi_a[s] \, u[C_1]
    \Bigr\rangle
\nonumber\\ && {}
    - \Bigl\langle
	{\tr\over \Nc} \>
	u[C_2]
    \Bigr\rangle
    \Bigl\langle
	{\tr\over \Nc} \>
	u[C_3] \, \phi_a[s]^\dagger \, u[C_4] \, \phi_a[s] \, u[C_1]
    \Bigr\rangle \,.
\label {eq:Oself parent}
\end {eqnarray}
In the daughter theory loop equation for ${\cal O}_\d$,
the analogous self-intersection terms coming from double variations
of $\phi^\j_a[s]$ or $\phi^{\j+\r_a}_a[s]$ are
\begin {eqnarray}
    \langle {\cal O}_\d \rangle_{\rm self-intersection}
    &=&
    - {1\over k^d} \sum_{\j\in T} \Bigl\langle
	{\tr\over N}
	u^{\j+\r_a}[C_1] \, \phi^{\j+\r_a}_a[s] \,
	u^{\j+2\r_a}[C_2] \, \phi^{\j+\r_a}_a[s]^\dagger \,
	u^{\j+\r_a}[C_3]
    \Bigr\rangle
    \Bigl\langle
	{\tr\over N}
	u^\j[C_4] \Bigr\rangle
\nonumber\\ &&
    - {1\over k^d} \sum_{\j\in T} \Bigl\langle
	{\tr\over N} \,
	u^{\j+2\r_a}[C_2]
    \Bigr\rangle
    \Bigl\langle
	{\tr\over N} \,
	u^{\j+\r_a}[C_3] \, \phi^\j_a[s]^\dagger \,
	u^\j[C_4] \, \phi^\j_a[s] \,
	u^{\j+\r_a}[C_1]
    \Bigr\rangle \,,
\nonumber\\
\label {eq:Oself daughter}
\end {eqnarray}
assuming that $\r_a \ne 0$.%
\footnote
    {
    If $\r_a\,{=}\,0$, then two additional self-intersection terms
    are generated which resemble the second and third terms
    in the parent theory result (\ref {eq:Oself parent}),
    namely
    $
	-\sum_\j
	\langle {\tr\over N} u^\j[C_1] \phi^\j_a[s] u^\j[C_2] \rangle
    $$
	\langle {\tr\over N} u^\j[C_3] \phi^\j_a[s]^\dagger u^\j[C_4] \rangle
	-\sum_\j
	\langle {\tr\over N} u^\j[C_2] \phi^\j_a[s]^\dagger u^\j[C_3] \rangle
    $$
	\langle {\tr\over N} u^\j[C_4] \phi^\j_a[s] u^\j[C_1] \rangle
    $.
    \label {fn:ra=0}
    }

Comparing the parent and daughter results,
there are two sources of ``mismatch''.
First and foremost, the two intersection terms
in the daughter theory result (\ref {eq:Oself daughter})
resemble the first and last terms in the parent
theory result (\ref {eq:Oself parent}),
but terms corresponding to the second and third terms of
the parent theory result are completely absent.
Second,
under the parent/daughter operator mapping 
every single-trace parent observable maps into
a sum over theory space of single-trace daughter observables.
Hence, each product of expectation values in the
parent self-intersection terms (\ref {eq:Oself parent})
should map onto a product of independent sums over theory space
of single expectation values.
Instead, the daughter theory intersection terms (\ref {eq:Oself daughter})
involve a single sum over theory space of a product of expectation values.

Dealing with the second point first,
note that the discrete $(Z_k)^d$ symmetry of theory space
guarantees that the expectation value of any daughter theory
operator involving terms at particular points in theory space will
coincide with the average of the expectation value over all points
in theory space --- {\em provided the $(Z_k)^d$ theory space symmetry
is not spontaneously broken in the daughter theory}.
Consequently, if the daughter theory is in a phase with
unbroken $(Z_k)^d$ symmetry, then
\begin {eqnarray}
    &&
    {1\over k^d} \sum_{\j\in T} \Bigl\langle
	{\tr\over N} \,
	u^{\j+\r_a}[C_1] \, \phi^{\j+\r_a}_a[s] \,
	u^{\j+2\r_a}[C_2] \, \phi^{\j+\r_a}_a[s]^\dagger \,
	u^{\j+\r_a}[C_3]
    \Bigr\rangle
    \Bigl\langle
	{\tr\over N} \,
	u^\j[C_4] \Bigr\rangle
\\ && {} =
    \biggl[
    {1\over k^d} \sum_{\j\in T} \Bigl\langle
	{\tr\over N} \,
	u^{\j+\r_a}[C_1] \, \phi^{\j+\r_a}_a[s] \,
	u^{\j+2\r_a}[C_2] \, \phi^{\j+\r_a}_a[s]^\dagger \,
	u^{\j+\r_a}[C_3]
    \Bigr\rangle
    \biggr]
    \biggl[
    {1\over k^d} \sum_{\j'\in T} \Bigl\langle
	{\tr\over N} \,
	u^{\j'}[C_4] \Bigr\rangle
    \biggr]
	\,,
\nonumber
\end {eqnarray}
{\em etc}.

To address the ``missing'' analogues of the second and third terms
in the parent theory result (\ref {eq:Oself parent}),
note that these terms involve expectation values of operators,
such as $\tr (\phi_a[s] \, u[C_2C_1] )$,
which are gauge invariant but are not invariant under $U(1)$ phase
rotations of the scalar field.
More specifically, the expectations in these terms are not
invariant under the $Z_k$ transformations $h_\alpha$ used
to define the orbifold projection,
under which $\phi_a$ acquires a phase of $e^{2\pi i r_\alpha/k}$.
(Unless $\r_a\,{=}\,0$, in which case these terms are invariant
and, as noted in footnote \ref {fn:ra=0}, analogues of these terms
do then appear in the daughter theory result.)
These $Z_k$ phase rotations are symmetries of the parent theory,
and these symmetries guarantee that expectation values
of operators transforming non-trivially (and irreducibly) under
these symmetries will vanish ---
{\em provided the non-gauge symmetries used to define
the orbifold projection are not spontaneously broken in the parent theory}.%
\footnote
    {
    Some readers may wonder whether these symmetry realization restrictions
    are necessary, since
    all symmetry non-invariant operators will have vanishing
    expectation values in a lattice measure containing no
    symmetry-breaking boundary conditions or perturbations ---
    regardless of the phase of the theory.
    Recall, however, that in the absence of symmetry breaking perturbations,
    spontaneous symmetry breaking is signaled by the breakdown of cluster
    decomposition in correlators of symmetry violating order parameters.
    Large $\Nc$ factorization, which we have used in deriving our
    loop equations, holds only in states which satisfy cluster decomposition.
    Therefore, in any phase of the theory which has spontaneous symmetry
    breaking, the lattice measure should tacitly be understood to include some
    perturbation which picks out a preferred equilibrium state satisfying
    cluster decomposition.
    }

To recap,
the second and third terms in the parent theory self-intersection result
(\ref {eq:Oself parent}) will vanish,
and the first and last terms will match the daughter theory
self-intersection result (\ref {eq:Oself daughter}),
provided the parent theory is in a phase which respects the
non-gauge symmetries used to construct the orbifold projection
and the daughter theory is in a phase which respects the
$(Z_k)^d$ translation symmetry of theory space.
Although these points have been illustrated with the
particular example of the operator (\ref {eq:O example}),
the conclusion is general:
the self-intersection terms in the loop equation of any
single-trace observable coincide in the parent and daughter theories
(under the parent/daughter operator mapping)
provided the orbifold and theory space symmetries are unbroken
in the parent and daughter theories, respectively.

The net result, under the assumption of appropriate unbroken symmetries,
is that the loop equations for arbitrary single-trace
observables in the daughter theory have exactly
the same form as in the parent theory, namely
\begin {eqnarray}
    \half \left(
	n_\l + V'[\left\langle\chi\right\rangle] \, n_{\rm s} +
	m \, n_{\rm f}
    \right) 
    \left\langle W_\d[C] \right\rangle
    &=&
    \sum_{\l \subset C} \sum_{p|\l \subset\partial p} \!\!
    \coeff 14 \betatilde^{(\d)}_{\l,p}
    \!\left\{
	\left\langle W_\d[(\overline {\partial p}) (\l \bar \l)^{-1} C]
	\right\rangle
	+
	s_{\l,p}
	\left\langle W_\d[(\partial p) C]\right\rangle
    \right\}\!
\nonumber\\ &-&
    \kern-5pt \sum_{\rm self-intersections} \kern-5pt
    I[\l] \,
    \left\langle W_\d[C'] \right\rangle
    \left\langle W_\d[C''] \right\rangle
\nonumber\\[3pt] &+&
    O(1/N^2) \,,
\label {eq:orbloop eq 4}
\end {eqnarray}
where the splitting coefficient $I[\l]$ is defined in
Eq.~(\ref {eq:ICCC}) and, as before,
$C = C'C''$ with loops
$C$, $C'$ and $C''$ all regarded as starting with link $\l$
for parallel traversals of the intersection link,
while $C = \l C' \bar\l C''$ for antiparallel traversals.%
\footnote
    {
    This geometric description of the
    self-intersection terms includes those terms which have
    just been argued to vanish in the parent theory.
    In the daughter theory, these terms correspond to
    splittings of the original loop $C$ (on the extended lattice $\bar\Lambda$)
    into subloops $C'$ and $C''$ each of which represent
    observables containing matter insertions whose $\r$ charges
    do not sum to zero.
    Such observables are not gauge-invariant in the daughter theory
    and their expectation values
    (in the gauge invariant measure of the theory)
    necessarily vanish.
    }

Implications of the equality of the above loop equations
between parent and daughter theories
(for single trace observables, suitably mapped between the two theories)
will be discussed in the next section.

\subsection {Multi-loop connected correlators}

Extending the derivation of loop equations to multi-loop connected
correlators is straightforward,
and proceeds in complete analogy with the parent theory treatment.
We only briefly sketch the two loop case.
Loop equations for two-loop connected correlators in the daughter theory
are generated by the identity
\begin{equation}
   \Big\langle \Delta \,
   \Big(\!\!
     \left(W_\d[C_1]-\langle W_\d[C_1]\rangle\right) 
     \left(W_\d[C_2]-\langle W_\d[C_2]\rangle\right) 
   \!\!\Big) \Big\rangle
   = 0 \,,
\end{equation}
with $\Delta$ given in Eq.~(\ref {eq:OrbDelta def}).
Evaluating this in the same fashion described previously,
using large $\Nc$ factorization
plus unbroken $(Z_k)^d$ translation invariance in theory space,
yields a loop equation which may be written in exactly the same form
as the previous parent theory result
(assuming unbroken orbifold projection symmetries in the parent),
namely
\begin {eqnarray}
    \half (
	n_\l \!&+&\! V'[\left\langle\chi\right\rangle] \> n_{\rm s} +
	m \> n_{\rm f}
    )\, 
    \dlangle W_\d[C_1] W_\d[C_2] \drangle
\nonumber\\[3pt]  &=&
    \Biggl[ \,
    \sum_{\l \subset C_1} \sum_{p|\l \subset\partial p}
    \coeff 14 \, \betatilde_{\l,p}
    \Bigl[
	\dlangle
	    W_\d[(\overline {\partial p}) (\l \bar \l)^{-1} C_1] \, W_\d[C_2]
	\drangle
	+
	s_{\l,p} \, \dlangle W_\d[(\partial p) C_1] \, W_\d[C_2] \drangle
    \Bigr]
\nonumber\\ 
    &-&
    \kern-5pt 
    \sum_{{\scriptstyle {\rm self-intersections} \atop \scriptstyle (C_1)}}
    \kern-5pt
    I[\l] \,
    \Bigl[
	\dlangle W_\d[C_1'] W_\d[C_2]\drangle
	\left\langle W_\d[C_1''] \right\rangle
	+ (C_1' \leftrightarrow C_1'')
    \Bigr]
 + (C_1 \leftrightarrow C_2)
 \Biggl]
\nonumber \\[2pt]
    &-&
    \kern -10pt
    \sum_{{\scriptstyle {\rm parallel\ gauge}
	\atop \scriptstyle {\rm mutual\ intersections}}
	\atop \scriptstyle \vphantom{\tilde C}(C_1,C_2)}
    \kern -10pt
	J[\l] \;
	\langle W_\d[C_1 C_2] \rangle
    \; -
    \kern -10pt
    \sum_{{\scriptstyle {\rm anti-parallel}
	\atop \scriptstyle {\rm mutual\ intersections}}
	\atop \scriptstyle \vphantom{\tilde C}(C_1,C_2)}
    \kern -10pt
	K[\l] \;
	\langle W_\d[C_1 (\l\bar\l)^{-1} C_2] \rangle 
\nonumber \\[2pt]
    &+&
    \kern -10pt
    \sum_{{\scriptstyle {\rm parallel\ gauge}
	\atop \scriptstyle {\rm mutual\ intersections}}
	\atop \scriptstyle (\widetilde C_1,C_2)}
    \kern -10pt
	J[\l] \;
	\langle W_\d[\widetilde C_1 C_2] \rangle
    \; +
    \kern -10pt
    \sum_{{\scriptstyle {\rm anti-parallel}
	\atop \scriptstyle {\rm mutual\ intersections}}
	\atop \scriptstyle (\widetilde C_1,C_2)}
    \kern -10pt
	K[\l] \;
	\langle W_\d[\widetilde C_1 (\l\bar\l)^{-1} C_2] \rangle 
\nonumber \\[5pt]
    &+&
    O(1/\Nc^2)
    \,.
\label {eq:2-loop eq daughter}
\end {eqnarray}
The coefficients $\betatilde_{\l,p}$, $s_{\l,p}$, $I[\l]$, $J[\l]$ and $K[\l]$
are all the same as defined previously
[{\em c.f.}, Eqs~(\ref{eq:beta tilde})--(\ref {eq:KCC})].
Therefore,
given the appropriate parent/daughter operator mapping
and the above assumptions concerning symmetry realizations,
the large $\Nc$ loop equations for both single trace expectation values
and two loop correlators coincide
between parent and daughter orbifold theories.%
\footnote
    {
    Although we have focused on bosonic observables up to now,
    this assertion about the correspondence between parent
    and daughter loop equations for two loop correlators is true
    for fermionic as well as bosonic loops;
    see Appendix~\ref{sec:Fermionic observables}.
    }

\section{Discussion}
\label{sec:Discussion}

The loop equations for single-trace expectation values
(\ref {eq:orbloop eq 4}) or correlators (\ref {eq:2-loop eq daughter})
may be solved iteratively (as described in section \ref {sec:Adjoint matter})
to generate expansions in powers of the plaquette weights
$\betatilde_{\l,p}^{(\d)}$,
or equivalently, double expansions in powers of the hopping parameter
$\kappa$ and the inverse 't Hooft coupling $\beta_p/\Nc$.
Consequently, the equality of large $\Nc$ loop equations
between parent and daughter theories
(under the appropriate operator mapping between the two theories)
implies that expectation values or correlators of
corresponding operators in the two theories have identical
strong coupling/large mass expansions (in the large $\Nc$ limit).
Standard methods for proving convergence of cluster expansions
\cite {OS, Seiler, LGY-lattice}
may be generalized without difficulty to theories with
product gauge groups such as the orbifold theories under consideration,
and show that the strong coupling/large mass expansions
in both parent and daughter theories have non-zero radii of convergence.
As a result, equality of the strong coupling/large mass expansions
immediately implies equality, within the radius of convergence,
of the exact expectation values or correlators themselves.
(This is why equality of strong coupling expansions is a much
stronger result than equality of weak coupling perturbation theory.)
And equality within the radius of convergence
immediately extends, via analytic continuation,
to exact equality throughout the portion of the
phase diagram in both theories which is continuously connected
to the strong coupling/large mass region.

As emphasized in the previous section, in order for the loop equations
of parent and daughter orbifold theories to coincide, the parent
theory must not spontaneously break the global symmetries used in
the orbifold projection, and the daughter theory must not spontaneously
break the discrete translation symmetry of theory space.
Within the strong coupling/large mass phase of either theory,
this is not an additional assumption;
the convergence of the strong coupling/large mass expansion
can easily be used to show that the $U(\Ns) \times U(\Nf)$ global
symmetry in the parent theory, and the $(Z_k)^d$ discrete symmetry
in the daughter theory, are unbroken within this phase.

It should be noted that the large-$\Nc$ equivalence between
parent and daughter orbifold theories which we have demonstrated
(within the strong coupling/large mass phase of both theories)
implies, in the large $\Nc$ limit, equality of the string tensions
of the two theories as well as equality of their spectrum of excitations
(within symmetry channels to which the parent/daughter operator mapping
applies, namely channels invariant under the global parent symmetries used
in the orbifold projection, and under theory space translations in the
daughter).
This merely reflects the fact that the string tension can be extracted from
large Wilson loops, and the mass spectrum from
the large distance behavior of correlators.

A further consequence of this large-$\Nc$ equivalence
is the existence of relations between correlation functions
of the daughter theory
which reflect symmetries that are present in the parent theory,
but absent in the daughter theory.
For example,
if differing $\r$-charges are assigned to the set of scalar fields
in a given orbifold projection,
or differing $\r$-charges are assigned to the set of fermions,
then the daughter theory will not be invariant under the
global $U(\Ns) \times U(\Nf)$ symmetry of the parent theory;
instead the daughter theory will only be invariant under
whatever subgroup of $U(\Ns) \times U(\Nf)$ preserves the $\r$-charge
assignments.
However, the large $\Nc$ equivalence with the parent theory means
that daughter theory correlation functions (in symmetry channels
to which the parent/daughter operator mapping applies) will satisfy
various $U(\Ns) \times U(\Nf)$ symmetry relations in the large $\Nc$ limit.
This implies that the particle spectrum of the large $\Nc$ daughter theory
must have degeneracies which do not follow from the symmetries
of the daughter theory --- but reflect projections of symmetry
relations in the parent theory
\cite {Strassler}.%
\footnote
    {
    A simple example is a parent theory containing two fermions,
    and a $U(2)$ flavor symmetry.
    A $Z_2$ orbifold projection with zero $r$-charge for one fermion
    and unit $r$-charge for the other yields a daughter theory with
    one set of adjoint fermions ($\chi^j$),
    one set of bifundamentals ($\psi^j$),
    and only a $U(1)\times U(1)$ flavor symmetry.
    The symmetry relations between correlators of the parent theory
    which survive projection to the daughter theory require, for example,
    that the two-point functions of
    $\sum_j \tr (\chi^j \chi^j)$ and $\sum_j \tr (\psi^j\psi^{j+1})$ coincide,
    implying degeneracy between the masses of single particle states containing
    two adjoint fermions and those with two bifundamentals.
    }

Extending our results to a wider class of theories should be straightforward,
but will be left to future work.
Possible extensions include consideration of more general orbifold projections
(such as cases where the projection-defining subgroup $H$ is non-Abelian),
inclusion of Yukawa couplings,
matter sectors with less flavor symmetry,
other gauge groups [$O(\Nc)$ or $Sp(\Nc)$],%
\footnote
    {
    It should be noted that $SU(\Nc)$ and $U(\Nc)$ gauge theories
    have coinciding large $\Nc$ limits;
    excluding (or including) the central $U(1)$ factor only affects
    subleading $1/\Nc^2$ suppressed contributions
    to either single-trace expectation values
    or connected multi-loop correlators.
    }
symmetric or antisymmetric tensor (instead of adjoint representation)
matter fields \cite{Armoni-Shifman}, and alternative fermion discretizations.

As our method of proof
(for the strong coupling/large mass phase of these lattice theories)
makes clear,
large $\Nc$ equivalence between parent and daughter orbifold
theories has nothing whatsoever to do with supersymmetry,
dimensionality, continuum limits, or large volume limits.
However, the extent to which this non-perturbative equivalence holds
outside the strong coupling/large mass phase is not yet clear.
Large $\Nc$ equivalence between parent and daughter theories clearly fails
to hold in any phase of the parent theory which spontaneously breaks
the particular global symmetries used in the orbifold projection,
as well as in any phase of the daughter theory which spontaneously breaks
the discrete theory space translation symmetry.
Such phases (when they exist) do not have equivalent loop equations.%
\footnote
    {
    The $Z_2$ orbifold of super-Yang-Mills theory
    compactified on $R^3 \times S^1$
    is an example of an orbifold theory with a phase in which
    the theory space translation symmetry is spontaneously broken
    (as shown by Tong \cite {Tong}),
    thereby invalidating the large $\Nc$ equivalence in this phase.
    }
But as long as these symmetries are not spontaneously broken,
then the loop equations of the two theories coincide.
The only way large $\Nc$ equivalence could fail in this circumstance
is if there are multiple physically acceptable solutions to the loop
equations, and the parent and daughter theories correspond to different solutions.
As mentioned in section \ref{sec:Pure gauge}, for simple models involving only
a few plaquettes, it is known that supplementing the loop equations by
trivial inequalities (reflecting unitarity of the gauge connection)
allows one to select the correct solution of the loop equations
on the weak coupling side of the Gross-Witten large $\Nc$ phase transition.
This may well be true more generally.
In any case,
extending our loop-equation based proof of large $\Nc$ equivalence
to other phases of lattice gauge theories, including weak coupling phases
with physical continuum limits, will require better
understanding of when (or if) parent and daughter theories
can correspond to different solutions of the same set of loop equations.

It is quite possible that a stronger version of our results,
valid beyond the strong coupling/large mass phase,
may be obtained by comparing the large $\Nc$ coherent state
variational actions \cite {LGY-largeN,BrownYaffe}
of the parent and daughter theories.
The minimum of this variational action yields
the free energy in the large $\Nc$ limit.
The loop equations for single trace operators are, in effect,
equations characterizing the location of stationary points of
this large $\Nc$ variational action,
but the value of the variational action itself is needed to determine
which stationary point describes the correct equilibrium state of the theory.
This is a topic for future work.

\acknowledgments

Josh Erlich, Herbert Neuberger, and Matt Strassler are thanked for
helpful comments.
The work of P.K. and L.G.Y. is supported, in part, by the U.S. Department
of Energy under Grant No.~DE-FG03-96ER40956;
the work of M.\"U. is supported by DOE grant  DE-FG03-00ER41132.

\newpage
\appendix

\section {Correlators of fermionic observables}
\label {sec:Fermionic observables}

Decorated Wilson loops containing an odd number of fermion insertions
do not correspond to closed loops on the extended lattice
$\bar\Lambda_{\rm f}$,
as it was defined in Section \ref {sec:encoding}.
Instead they correspond to open contours in $\bar\Lambda_{\rm f}$
whose endpoints are $Z_2$ partners of each other.
This is a perfectly consistent representation of single-trace
fermionic observables,
although it has the drawback of involving a distinguished
starting (and ending) site on the loop.
In other words, this representation does not make
trace cyclicity manifest.

The previous construction of the extended lattice $\bar\Lambda_{\rm f}$ was
dictated by the desire to represent, geometrically,
bosonic observables containing an even number of fermion insertions.
Such observables only satisfy trace cyclicity up to a sign, which
is why it was necessary for the extended lattice $\bar\Lambda_{\rm f}$ 
to involve a doubling of sites in the original lattice $\Lambda$
(so that a loop and its $Z_2$ `mirror' will represent the same observable,
but with opposite overall signs).
In contrast,
fermionic single-trace observables {\em do} satisfy
trace cyclicity, without any minus signs
(because moving a fermion insertion from one end to the other now involves
an even number of fermionic transpositions).
So for fermionic observables, an alternate, and simpler, geometric
encoding is to represent these observables as ordinary closed loops
on the smaller extended lattice $\bar\Lambda_{\rm f}/Z_2$, in which
all $Z_2$ partner sites $s$ and $\tilde s$ in $\bar\Lambda_{\rm f}$
are now identified.
(Hence the result looks just like the minimally extended lattice
$\bar\Lambda_{\rm s}$ for scalars).

Using either representation, one may generalize the previous treatment
of loop equations for decorated loops,
described in Section \ref {sec:LE-parent},
to the case of fermionic loops.
All expectation values of fermionic loops vanish,%
\footnote
    {
    One might wonder if it is ever possible to break spontaneously
    the $Z_2$ symmetry [often called $(-1)^F$]
    which distinguishes fermions from bosons.
    We will ignore this perverse possibility.
    }
so the only relevant fermionic loop equations are those
for connected correlators involving an even number of
fermionic loops.
For two-loop correlators,
the result may be written in the form (\ref {eq:2-loop eq 2})
previously derived for bosonic loops, except that
there is no need to include the mutual intersection terms involving
$Z_2$ shifted loops if the $\bar\Lambda_{\rm f}/Z_2$ representation
is used.%
\footnote
    {
    More precisely, one should ignore the $-1$ factors in $I[\l]$ and $J[\l]$
    associated with intersection links starting on $\widetilde\Lambda$,
    and redefine $K[\l]$ to be $-1$ for gauge, scalar,
    or backward-directed fermion ($\psi$) links,
    and $+1$ for forward-directed fermion ($\bar\psi$) links.
    In the self intersection and mutual intersection terms,
    loops appearing in single trace expectation values
    (which correspond to bosonic observables)
    should be ``lifted'' to the original extended lattice
    $\bar\Lambda_{\rm f}$ by regarding the intersection link $\l$
    as starting from a site in the physical lattice $\Lambda$.
    }

Considering correlators of fermionic operators instead of
bosonic operators makes no difference as far the equivalence
between parent and daughter orbifold loop equations is concerned.
This presumes, of course, that gauge invariant fermionic operators exist
in the daughter theory; this will depend on the chosen $\r$-charge assignments.
If the chosen $\r$-charges do allow
fermionic single trace operators in the daughter theory,
then the loop equations for connected correlators
of fermionic loops coincide (under the parent/daughter operator mapping)
under the same conditions needed for coinciding bosonic correlators ---
the parent theory must not spontaneously break the symmetries
used in the orbifold projection and the daughter theory must not
spontaneously break its theory space translation symmetry.

\newpage

\section {Iterative solution of loop equations}
\label {sec:Iteration}

It is instructive to see how the physics of confinement
emerges directly from the loop equations in the strong coupling
limit of a lattice gauge theory.%
\footnote
    {
    There are arguably more direct methods
    for generating the strong coupling expansion of lattice gauge
    theories \cite {Wilson,Seiler}.
    The point of this appendix is merely to illustrate
    how the minimal set of loop equations (\ref {eq:loop eq 2})
    suffice for extracting this physics.
    }
For simplicity, consider a pure $U(\Nc)$ gauge theory formulated on a
simple cubic lattice, and take the coupling
$\betatilde\equiv\beta_p/\Nc$ to be the same for all plaquettes.
To generate the strong coupling (small $\betatilde$)
expansion of Wilson loop expectation values,
one may imagine assembling all possible loops into an (infinitely) long
vector, and then repeatedly iterating the
the loop equations (\ref {eq:loop eq 2}) for this vector,
starting with zero expectation values for all loops
except the trivial loop $\langle 1\rangle = 1$.
After a single iteration, one finds that the only loops with $O(\betatilde)$
expectations are elementary plaquettes,
\begin{equation}
     \langle\, \framebox[0.4cm]{\phantom{x}}\,\rangle =
     \frac{\betatilde}{2} \, \langle 1\rangle + \hbox {(higher-order)} \,.
\label{eq:iter-1}
\end{equation}
On the right-hand side of (\ref{eq:iter-1}), the
trivial loop appears in the deformation terms,
which remove a plaquette from the original loop.
Other deformation terms, which attach a plaquette to the original loop,
vanish at this order in the iteration (as well as the next),
and lead to higher order corrections [of order $\betatilde^3$].
At second order in the iteration, one finds $O(\betatilde^2)$
expectation values for two-plaquette loops.
For example,
\begin{equation}
     \langle\, \framebox[0.8cm]{\vphantom{x}}\,\rangle =
     \left( \frac{\smash{\betatilde}\vphantom\beta}{2} \right)^2 +
     \hbox{(higher-order)} \,.
\end{equation}
One may see directly from the loop equations that
every plaquette deformation of a loop is associated with one factor
of $\betatilde$.
Consequently, for a general loop of area $A$
({\em i.e.}, a loop whose minimal spanning surface contains $A$ plaquettes),
it is easy to see that one must iterate the loop equations $A$ times
before generating a non-zero contribution, so that
$\langle W[C] \rangle = O(\betatilde^A)$.
Determining the coefficient is easy once one realizes that
the number of deformation terms leading to decrease in the area
of a loop [in the minimal loop equations (\ref {eq:loop eq 2})]
precisely equals the number of links forming the loop.
These terms give identical contributions (at leading order)
and, in effect, cancel the factor of the loop perimeter on the left side of the
loop equations.
Consequently, one finds confining area-law behavior,
\begin{equation}
     \langle W[C]\rangle = 
     \left( \frac{\smash{\betatilde}\vphantom\beta}{2} \right)^A
     + \hbox{(higher-order)} \,,
\end{equation}
or $\langle W[C] \rangle \sim e^{-\sigma A}$ with a string tension
(in lattice units)
\begin{equation}
     \sigma = \ln\frac{2}{\betatilde} \,,
\end{equation}
up to sub-leading corrections.
With a bit more effort,
one may show that corrections to the string tension
are $O(\betatilde^4)$.

It is straightforward to repeat
the analysis when adjoint matter fields are present.
If both $\betatilde$ and $\kappa/m$ are small
(corresponding to strong coupling and large mass),
then a similar iteration of the loop equations shows that
Wilson loops still exhibit area law behavior
(which would not be the case, of course, with
fundamental representation matter fields).
Dynamical adjoint matter fields only generate
contributions to the string tension which are suppressed by at least $\kappa^6$,
\begin{equation}
     \sigma = \ln\Bigl(\frac{2}{\betatilde}\Bigr)
		+ O(\betatilde^4)
		+ O\Bigl(\frac {\kappa^6} {m^6}\Bigr)
		\,.
\end{equation}

%
%
\newsavebox{\twoplaq}
\savebox{\twoplaq}
    {
    \setlength{\unitlength}{1mm}
    \begin{picture}(17,4)(-1.5,2)
	    \drawline(0,0)(0,4)
	    \drawline(0,0)(2,2)
	    \drawline(2,2)(2,6)
	    \drawline(2,6)(0,4)
	    \drawline(10,0)(10,4)
	    \drawline(10,0)(12,2)
	    \drawline(12,2)(12,6)
	    \drawline(12,6)(10,4)
	    \put(1,0){\vector(1,0){8.5}}
	    \put(9,0){\vector(-1,0){8.5}}
	    \put(5,1){\scriptsize$L$}
	    \put(-4,2){\Large$\dlangle$}
	    \put(13,2){\Large$\drangle$}
    \end{picture}
    }

A similar iterative approach may be applied to the loop equations
(\ref {eq:2-loop eq})
for connected correlators of Wilson loops.
Consider, for example, the correlator of two elementary plaquettes
separated by a lattice distance $L$, $\usebox{\twoplaq}$.
Iterating the loop equations (\ref {eq:2-loop eq}),
starting with all two-loop connected correlators equal to zero,
one may easily see that non-zero contributions will only arise
after some sequence of plaquette deformations acting on
one or the other plaquette (or both) causes the deformed loops
to have a mutual intersection.
Since each plaquette deformation costs a factor of $\betatilde$,
the leading contribution must involve a power of $\betatilde$
which is proportional to $L$.
Determining the correct power (directly from the loop equations)
is a bit tricky.
After $L{+}1$ iterations of the loop equations, one first finds
mutual intersection terms of the form
\begin {equation}
    \usebox{\twoplaq} \sim \betatilde^L
    \setlength{\unitlength}{1mm}
    \begin{picture}(18.5,4)(-3.5,2)
	    \drawline(0,0)(0,4)
	    \drawline(0,0)(2,2)
	    \drawline(0,4)(10,4)
	    \drawline(2,2)(2,6)
	    \drawline(10,0)(10,4)
	    \drawline(10,0)(12,2)
	    \drawline(12,6)(2,6)
	    \drawline(12,2)(12,6)
	    \put(1,0){\vector(1,0){8.5}}
	    \put(9,0){\vector(-1,0){8.5}}
	    \put(5,1){\scriptsize$L$}
	    \put(-3,2){\Large$\langle$}
	    \put(13,2){\Large$\rangle$}
    \end{picture}
    + \hbox {(other terms)}.
\end {equation}
The Wilson loop on the right has an $O(\betatilde^{L+2})$
expectation value, so one might expect the two plaquette
connected correlator to be $O(\betatilde^{2L+2})$.
However, there are cancellations between deformation and
mutual intersection terms, which eliminate all contributions
below order $\betatilde^{4L}$.
A more careful analysis shows that the first non-zero contribution
comes from deformations which build a ``tube'' between the two initial
plaquettes, so that%
\footnote
    {
    This result is valid in three or more dimensions.
    In two dimensions, the plaquette-plaquette correlation function
    vanishes identically (except when the plaquettes coincide).
    One can see this from the fact that in
    $d{=}2$, a suitable choice of gauge allows one to rewrite the
    integral over link variables as an integral over independent
    plaquette variables.
    The resulting partition function factorizes into a product of
    single plaquette contributions, and correlations between
    different plaquettes are absent.
    }
\begin {equation}
    \usebox{\twoplaq}
    =
     \left( \frac{\smash{\betatilde}\vphantom\beta}{2} \right)^{4L}
     \times [1 + O(\betatilde^2)] \,.
\end {equation}
Consequently, the correlator falls exponentially
with distance, $\usebox{\twoplaq} \sim e^{-\mu L}$, with
a mass gap $\mu$ (equal to the lightest glueball mass) given by
\begin{equation}
     \mu = 4 \, \ln \frac {2}{\betatilde} \,,
\end{equation}
up to sub-leading corrections [which turn out to be $O(\betatilde^4)$].
As with the string tension,
inclusion of adjoint matter fields only produces at most
$O(\kappa^6)$ sub-leading corrections to the mass gap $\mu$.

\newpage
\sloppy

\end {document}